\documentclass[prd,preprintnumbers,nofootinbib]{revtex4} 
\usepackage{graphicx} 
\usepackage{amsmath}
\usepackage{amsfonts,amsbsy}
\usepackage{amssymb}
\usepackage{appendix}

\def\be{\begin{equation}}
\def\ee{\end{equation}}
\def\bea{\begin{eqnarray}}
\def\eea{\end{eqnarray}}

\def\Tr{{\rm tr}}
\def\eq#1{{Eq.~(\ref{#1})}}
\def\fig#1{{Fig.~\ref{#1}}}

\begin{document}

\title{\bf Prompt photon production and photon-hadron correlations at RHIC and the LHC from the  Color Glass Condensate}


\author{Jamal Jalilian-Marian$^{1,2}$ and Amir H. Rezaeian$^{3}$}
\affiliation{
$^1$ Department of Natural Sciences, Baruch College, CUNY,
17 Lexington Avenue, New York, NY 10010, USA\\
$^2$ The Graduate School and University Center, City
  University of New York, 365 Fifth Avenue, New York, NY 10016, USA\\
$^3$Departamento de F\'\i sica, Universidad T\'ecnica
Federico Santa Mar\'\i a, Avda. Espa\~na 1680,
Casilla 110-V, Valparaiso, Chile }

\begin{abstract}
We investigate inclusive prompt photon and semi-inclusive prompt
photon-hadron production in high energy proton-nucleus
collisions using the Color Glass Condensate (CGC) formalism which 
incorporates non-linear dynamics of gluon saturation at small $x$ via Balitsky-Kovchegov equation with running coupling. 
For inclusive prompt photon production, we rewrite the cross-section 
in terms of direct and fragmentation contributions and show that the 
direct photon (and isolated prompt photon) production is more sensitive to 
gluon saturation effects. We then analyze azimuthal correlations in photon-hadron production in high 
energy proton-nucleus collisions and obtain a strong suppression of 
the away-side peak in photon-hadron correlations at forward rapidities, similar to the observed mono-jet 
production in deuteron-gold collisions at forward rapidity at RHIC.  
We make predictions for the nuclear modification factor $R_{p(d)A}$ and photon-hadron azimuthal 
correlations in proton(deuteron)-nucleus collisions at RHIC and the LHC at various rapidities.
\end{abstract}

\maketitle

\section{Introduction}

The Color Glass Condensate (CGC) formalism has been successfully applied to many processes in high energy collisions involving at least one hadron or nucleus in the initial state. Examples are structure functions (inclusive and diffractive) in Deeply Inelastic Scattering of electrons on protons or nuclei, and particle production in proton-proton, proton-nucleus and nucleus-nucleus collisions, for a recent review see Ref.\,\cite{cgc-review1}. The predicted suppression of $R_{dA}$ for the single inclusive hadron production in deuteron-nucleus (dA) collisions as well as the disappearance of the away side peak in di-hadron angular correlations in the forward rapidity region of RHIC~\cite{for-sup,cor-sup} are two of most robust predictions of the formalism which have been confirmed \cite{exp}, see also Refs.\,\cite{j-c,me-jamal1,posdiction}. The CGC formalism has been also successful in providing predictions for the first LHC data \cite{lhcd} in proton-proton (pp) and nucleus-nucleus collisions \cite{m1,j1,lhc-o}. Nevertheless, there are more recent, alternative phenomenological approaches which combine nuclear shadowing, transverse momentum broadening and cold matter energy loss to describe the RHIC data~\cite{pqcd-models,pqcd-models-2}. Therefore, one needs to consider other observables which may help clarify the underlying dynamics of forward rapidity particle production at small $x$. 

Inclusive prompt photon production~\cite{pho-cgc} and prompt photon-hadron angular correlations~\cite{pho-had-cor-cgc} in the forward rapidity region are two such examples. Furthermore, there are advantages to studying 
prompt photon production as compared to hadron production. It is theoretically cleaner; one avoids the difficulties involved with description of hadronization of final state quarks and gluons, usually described by fragmentation functions valid at high transverse momentum. Also, one does not have to worry about possible initial state-final state interference effects which may be present for hadron production. In case of photon-hadron vs. di-hadron angular correlations, again the underlying theoretical understanding is more robust. Unlike di-hadron
correlations which involve higher number of Wilson lines~\cite{n-point}, photon-hadron correlations depend only on the dipole cross section properties of which are well understood.

Both processes have been investigated previously, albeit not in detail and only in a  limited kinematic range \cite{pho-cgc,pho-had-cor-cgc}, see also Refs.\,\cite{boris,me2-pho}. In this work, we extend the existing results for inclusive prompt photon production by clearly separating the contribution of direct and fragmentation photons. We show that direct photons are more sensitive to gluon saturation effects in the kinematics regions considered. We then investigate the dependence of prompt photon-hadron azimuthal angular correlations on high gluon density effects and show that gluon saturation effects lead to disappearance of the away side peak. The effect is very similar to the disappearance of the away side peak in di-hadron correlations observed in the forward rapidity region of RHIC in dA collisions ~\cite{alba-mar}. Therefore, a measurement of this correlation at RHIC and the LHC would greatly help to clarify the role of CGC in the dynamics of particle production at high energy.

The advantage of the CGC formalism over the more phenomenological models is that the cross section for many of these processes have the same common ingredient~\cite{cgc-review1,dhj,m-b}, the dipole total cross section; the imaginary part of the forward scattering amplitude of a quark-antiquark dipole on a proton or nucleus target. Its rapidity (energy) dependence is governed by the B-JIMWLK/BK evolutions equations~\cite{jimwlk,bk} and is pretty well-understood. The most recent
advances in our understanding of the rapidity dependence of the dipole cross section include the running coupling constant corrections and the full Next-to-Leading Order corrections~\cite{nlo}. The only input is the dipole profile (dependence on the dipole size $r_t$) at the initial rapidity $y_0$ which is modeled, usually motivated by the McLerran-Venugopalan (MV) model~\cite{mv}. The sensitivity to this initial condition is expected to go away at very large rapidities, see Sec. III and Ref.\,\cite{me-jamal1}. 

This paper is organized as follows; we consider prompt photon-hadron production cross section in section IIA  and inclusive prompt photon production in section IIB where we describe how to separate the contribution of direct and fragmentation photons. We then present our detailed numerical results and predictions at kinematics appropriate for RHIC and the LHC experiments in section III. We summarize our results in section IV.

\section{Theoretical framework}
\subsection{Semi-inclusive Photon-hadron production in proton-nucleus (pA) collisions} 

The cross section for production of a quark and a prompt photon with $4$-momenta $l$ and $k$ respectively (both on-shell) in scattering of a on-shell quark with $4$-momentum $p$ on a target (either proton or nucleus) in the
CGC formalism has been calculated in \cite{pho-cgc} and is given by 
\be d\, \sigma = \frac{e^2\,
  e_q^2}{2}\frac{d^3 k}{(2\pi)^3\,2 k^-} \frac{d^3 l}{(2\pi)^3\, 2
  l^-} \frac{1}{2 p^-}(2\pi)\, \delta (p^- - l^- -k^-) \, \Tr_D\,
     [\cdots]\,d^2 \vec{b_t}\, d^2 \vec{r_t}\, e^{i (\vec{l_t} + \vec{k_t})\cdot
       \vec{r_t}} \, N_F (b_t, r_t, x_g),
\label{cs_gen}
\ee
 where $Tr_D\, [\cdots]$ is given by
\be
\Tr_D\, [\cdots] = 8\, [(p^-)^2 + (l^-)^2] \bigg[\frac{p\cdot l}{p\cdot k\, l\cdot k} + 
\frac{1}{l\cdot k} - \frac{1}{p\cdot k} \bigg],
\ee
which, after using the explicit forms of the momenta in the expression for the trace, can be 
written as,  
\bea
&&{d\sigma^{q(p)\, T \rightarrow q(l)\,\gamma(k)\, X}
\over d^2\vec{b_t}\, dk_t^2\, dl_t^2\, dy_{\gamma}\, dy_l\, d\theta} =
{e_q^2\, \alpha_{em} \over \sqrt{2}(2\pi)^3} \, 
{k^-\over  k_t^2 \sqrt{S}} \,
{1 + ({l^-\over p^-})^2 \over
[k^- \, \vec{l_t} - l^- \vec{k_t}]^2}\nonumber \\
&&\delta [x_q - {l_t \over \sqrt{S}} e^{y_l} - {k_t \over \sqrt{S}} e^{y_{\gamma}} ] \,
\bigg[ 2 l^- k^-\, \vec{l_t} \cdot \vec{k_t} + k^- (p^- -k^-)\, l_t^2 + l^- (p^- -l^-)\, k_t^2 \bigg] 
\nonumber \\
&&\int d^2 \vec{r_t} \, e^{i (\vec{l_t} + \vec{k_t})\cdot\vec{r_t}}   \, N_F (b_t, r_t, x_g) ,
\label{cs}
\eea
where the symbol $T$ stands for a proton $p$ or a nucleus $A$ target,  $\sqrt{S}$ is the nucleon-nucleon center of mass energy and $x_q$ is the ratio
of the incoming quark to nucleon energies such that $p^-=x_q \, \sqrt{S/2}$. The 
outgoing photon and quark rapidities are defined via $k^-={k_t \over \sqrt{2}} e^{y_{\gamma}}$
and $l^-={l_t \over \sqrt{2}} e^{y_l}$ whereas $\Delta \theta$ is angle between the final state
quark and photon, $cos(\Delta \theta) \equiv {\vec{l}_t \cdot \vec{k}_t \over  l_t k_t}$. We note 
that this cross section was first computed in~\cite{boris} in coordinate space using the dipole formalism, 
the result of which agrees with the expression in \eq{cs} after Fourier transforming to momentum space.

The imaginary part of of (quark-antiquark) dipole-target  forward scattering amplitude $N_F (b_t, r_t, x_g)$ satisfies the B-JIMWLK
equation and has all the multiple scattering and small $x$ evolution effects encoded. 
It 
is defined as 
\be
N_F(b_t,r_t,x_g) = {1\over N_c} \, < Tr [1 - V^{\dagger} (x_t) V (y_t) ] >,
\label{cs_def}
\ee
where $N_c$ is the number of color. The vector $\vec{b_t}\equiv (\vec{x_t} + \vec{y_t})/2$ is the impact parameter of the dipole from the target and $\vec{r_t}\equiv \vec{x_t} - \vec{y_t}$ denotes the dipole transverse vector. 
The matrix $V (y_t)$ is a unitary matrix in fundamental representation of $SU(N_c)$ containing the interactions of a quark and the colored glass condensate target. The dipole scattering probability 
depends on Bjorken $x_g$ via the B-JIMWLK renormalization group equations. In the present case,
it is related to the prompt photon and final state quark rapidities and transverse momenta via
\be
x_g = {1\over \sqrt{S}}[k_t e^{-y_{\gamma}} + l_t e^{-y_l}]\, .
\label{x_g}
\ee

In order to relate the above partonic production cross-section to 
proton (deuteron)-target collisions, one needs to convolute the 
partonic cross-section in \eq{cs} with the quark and
antiquark distribution functions of a proton (deuteron) and the quark-hadron 
fragmentation function:
\begin{eqnarray}\label{qh-f}
\frac{d\sigma^{p\, T \rightarrow \gamma (k)\, h (q)\, X}}{d^2\vec{b_t} \, dk^2_t\, dq^2_t \,  
d\eta_{\gamma}\, d\eta_{h}d\theta}&=& \int^1_{z_{f}^{min}} \frac{dz_f}{z_f^2} \, 
 \int\, dx_q\,
f (x_q,Q^2) \frac{d\sigma^{q\, T \rightarrow \gamma \, q \, X}} {d^2\vec{b_t}\, dk^2_t\,dl^2_t\, 
d\eta_{\gamma}\, d\eta_{h}\, d\theta} D_{h/q}(z_f,Q^2),
\end{eqnarray}
where $q_t$ is the transverse momentum of the produced hadron, and $f(x_q,Q^2)$ is the 
parton distribution function (PDF) of the incoming proton (deuteron) which depends on 
the light-cone momentum fraction $x_q$ and the hard scale $Q$. A summation over the 
quark and antiquark flavors in the above expression should be understood.
The function $D_{h/q}(z_f,Q)$ is the quark-hadron fragmentation function (FF) 
where $z_f$ is the ratio of energies of the produced hadron and quark~\footnote{ Since 
produced hadrons are assumed to be massless, we make no distinction between the
rapidity of a quark and the hadron to which it fragments. Moreover, for massless hadrons, rapidity $y$ and pseudo-rapidity $\eta$ is the same.}. Note that due to the assumption
of collinear fragmentation of a quark into a hadron, the angle $\Delta \theta$ is now the
angle between the produced photon and hadron.

The light-cone momentum fraction $x_q, x_{\bar q}, x_g$ are related to the transverse momenta and
rapidities of the produced hadron and prompt photon via (details are given in the appendix)
\begin{eqnarray}\label{qh-k}
x_q&=&x_{\bar{q}}=\frac{1}{\sqrt{S}}\left(k_t\, e^{\eta_{\gamma}}+\frac{q_t}{z_f}\, e^{\eta_{h}}\right),\nonumber\\
x_g&=&\frac{1}{\sqrt{S}}\left(k_t\, e^{-\eta_{\gamma}}+ \frac{q_t}{z_f}\, e^{-\eta_{h}}\right),\nonumber\\
z_f&=&q_t/l_t \hspace{1 cm} \text{with}~~~~~ z_{f}^{min}=\frac{q_t}{\sqrt{S}}
\left(\frac{e^{\eta_h}}
{1 - {k_t\over \sqrt{S}}\, e^{\eta_{\gamma}}}\, 
\right).\label{z_f}\label{ki1}\
\end{eqnarray}

\subsection{Single inclusive prompt photon production in proton-nuclear collisions} 

The single inclusive prompt photon cross section can be readily obtained from
\eq{cs_gen} by integrating over the momenta of the final state quark. Integration
over the quark energy $l^-$ is trivially done by using the delta function and leads 
to (after shifting $\vec{l_t} \rightarrow \vec{l_t} - \vec{k_t}$) , 
\begin{eqnarray}\label{pho}
\frac{d\sigma^{q (p) T \rightarrow \gamma (k) \, X}}{d^2\vec{b_t} dk^2_t d\eta_{\gamma}}&=& 
\frac{e_q^2 \alpha_{em}}{(2\pi)^3} z^2[1+(1-z)^2]
\frac{1}{k^2_t} \int d^2 \vec{r_t} \,d^2 \vec{l_t} 
\frac{l_t^2}{[z\, \vec l_t - \vec k_t ]^2}
\, e^{i \vec{l_t}\cdot \vec{r_t}}   \, N_F (b_t, r_t, x_g), 
\end{eqnarray}
where $z \equiv k^-/p^-$ denotes the fraction of the projectile quark energy $p^-$ carried by
the photon and $d\eta_\gamma \equiv \frac{d z}{z}$. Various limits of this expression have 
been studied in \cite{pho-cgc} where it was shown that in the limit where photon has a large transverse momentum $k_t \gg z\, l_t$ such that the collinear singularity is suppressed, one recovers the 
LO pQCD result for direct photon production process $q\, g \rightarrow q\, \gamma$ 
convoluted with the unintegrated gluon distribution function of the target. On the other hand,
if one performs the $l_t$ integration above without any restriction, one recovers  
the LO pQCD expression for quark-photon fragmentation function convoluted with 
dipole scattering probability. In the limit where one can ignore multiple scattering of
the quark on the target ("leading twist" kinematics), this expression reduces to the 
pQCD one describing LO production of fragmentation photons. It is therefore useful to 
explicitly separate the contribution of fragmentation photons from that of the direct photons.
To this end, we rewrite \eq{pho} as
\begin{eqnarray}\label{pho1}
\frac{d\sigma^{q (p)\, T \rightarrow \gamma (k) \, X}}{d^2\vec{b_t} d^2\vec{k_t} d\eta_{\gamma}}&=& 
\frac{e_q^2 \alpha_{em}}{\pi (2\pi)^3} z^2[1 + (1-z)^2]
\frac{1}{k^2_t} 
\int d^2 \vec{l_t}\, l_t^2\Big[\frac{1}{[z \vec l_t -\vec k_t]^2}-\frac{1}{k_t^2}\Big]
N_F(x_g,b_t,l_t), \nonumber\\
&+& \frac{e_q^2 \alpha_{em}}{\pi (2\pi)^3} z^2 [1 + (1 - z)^2]\frac{1}{k^4_t}\int d^2 \vec{l_t} \, l_t^2
N_F(x_g,b_t,l_t), 
\end{eqnarray}
where we have added and subtracted the second term.  Notice that we use the same notation for coordinate representation of the forward dipole-target scattering amplitude $N_F$ and its two-dimensional Fourier transform.  The second term in this expression 
describes production of direct photons whereas the first term gives the contribution of
fragmentation photons. In order to see this more explicitly, we let $\vec{l_t} \to \vec{l_t} + \frac{\vec{k_t}}{z}$ 
in the first term and keep the most divergent piece of the $l_t$ integral to get 
\begin{eqnarray}\label{pho2}
\frac{d\sigma^{q (p)\, T \rightarrow \gamma (k) \, X}}{d^2 \vec{b_t} d^2 \vec{k_t} d\eta_{\gamma}}&=&
\frac{d\sigma^{\text{Fragmentation}}}{d^2 \vec{b_t} d^2 \vec{k_t} d\eta_{\gamma}}+\frac{d\sigma^{\text{Direct}}}{d^2 \vec{b_t} d^2 \vec{k_t} d\eta_{\gamma}} 
\nonumber\\
&=&\frac{1}{(2\pi)^2}\frac{1}{z}\, D_{\gamma/q}(z,k_t^2)\, 
N_F(x_g,b_t,k_t/z) + 
 \frac{e_q^2 \alpha_{em}}{\pi (2\pi)^3}z^2[1+(1 - z)^2]\frac{1}{k^4_t}
\int^{k^2_t}d^2\vec{l_t}\,l_t^2\, 
N_F(\bar{x}_g,b_t,l_t) 
\end{eqnarray}
where  $D_{\gamma/q}(z,k_t^2)$ is the leading order quark-photon fragmentation function \cite{own}, 
\begin{equation}\label{pho3}
D_{\gamma/q}(z,Q^2)=\frac{e_q^2 \alpha_{em}}{2\pi}\frac{1 + (1 - z)^2}{z}\ln{Q^2/\Lambda^2}.
\end{equation}
\eq{pho2} is new and is our main result for single inclusive prompt photon production which includes  
contribution of both fragmentation (first term) and direct (second term) photons. 
In order to ensure that the divergence present in \eq{pho} is properly 
removed, one needs to regulate it self-consistently. Here we have done this 
separation by imposing a hard cutoff which would result in a mismatch between
the finite corrections to our results and those that are included in parameterizations
of photon fragmentation function, for example, using the $\overline{MS}$ scheme. However 
this mismatch is a higher order effect in the coupling constant and is therefore expected to be 
parametrically small. It should be noted that the separation between the direct and fragmentation 
contributions depends on the hard scale, chosen to be the photon transverse momentum, 
which is already well-known in pQCD.

\eq{pho2} exhibits some interesting features; the dipole scattering probability $N_F$
is probed at $k_t/z$ (where $k_t$ is the external momentum) in case of fragmentation 
photons whereas it depends on the internal momentum $l_t$ in case of direct photons. 
Furthermore, in case of direct photons, the integrand is peaked at values of transverse 
momenta $l_t \sim Q_s$. This means that fragmentation photons should be much less sensitive 
to high gluon density effects than direct photons since they probe the target structure at 
higher transverse momenta. This will be verified numerically in the following sections.  

In order to relate the partonic cross-section given by \eq{pho2} to photon production in 
deuteron (proton)-nucleus collisions, we convolute \eq{pho2} with quark and antiquark 
distribution functions of the projectile deuteron (or proton), 
 \begin{equation}\label{pho4}
\frac{d\sigma^{p\, T \rightarrow \gamma (k) \, X}}{d^2\vec{b_t} d^2\vec{k_t} d\eta_{\gamma}}=  
\int_{x_q^{min}}^1 d x_q [f_q(x_q,k_t^2)+ f_{\bar{q}}(x_{\bar q},k_t^2)] 
\frac{d\sigma^{q (p) \, T \rightarrow \gamma (k) \, X}}{d^2 \vec{b_t} d^2\vec{ k_t} d\eta_{\gamma}},
\end{equation}
where a summation over different flavors is understood. Equations (\ref{pho2},\ref{pho4}) are 
our final results for the single inclusive prompt photon production. The light-cone fraction 
variables $x_g,\bar{x}_g,z$ in Eq.~(\ref{pho2},\ref{pho4})  are 
defined as follows,
\begin{eqnarray}\label{pho5}
x_g&=& \frac{k_t^2}{z^2\, x_q\, S} = x_q \, e^{-2\, \eta_\gamma} \\
\bar{x}_g &=& \frac{1}{x_q\, S} \left[{k_t^2\over z} + \frac{(l_t-k_t)^2}{1-z}\right]\approx \frac{1}{x_q\, S} {k_t^2 \over z (1-z)},\label{xg-a}\\
z&\equiv& \frac{k^-}{p^-} = \frac{k_t}{x_q\, \sqrt{S}}e^{\eta_{\gamma}} = \frac{x_q^{min}}{x_q} 
\hspace{1 cm} \text{with}~~~~~ x_q^{min}=z_{min}=\frac{k_t}{\sqrt{S}}e^{\eta_{\gamma}}. \
\end{eqnarray}
where in \eq{xg-a} the right hand-side approximation is valid if $l_t\ll k_t$. 
Notice that since now $\bar{x}_g$ depends on the angle 
between $l_t$ and $k_t$, the integral over the angle in \eq{pho2} is
not more trivial and can be done numerically. One should also note that the
light-cone fraction variables defined above for the inclusive prompt photon cross-section Eqs.~(\ref{pho2},\ref{pho4}) are different from the corresponding semi-inclusive hadron-photon cross-section
Eqs.~(\ref{cs},\ref{qh-f}) defined in Eqs.~(\ref{ki1}), see the appendix for the derivation.

\section{Numerical results and predictions} 
The forward dipole-target scattering amplitude appears in both semi-inclusive photon-hadron and inclusive prompt photon cross-section Eq.~(\ref{qh-f},\ref{pho4}) and incorporates small-x dynamics which can be 
computed via first principle non-linear B-JIMWLK equations \cite{jimwlk} in the CGC formalism. In the large $N_c$ limit, 
the coupled B-JIMWLK equations are simplified to the Balitsky-Kovchegov (BK) equation \cite{bk}, a closed-form equation for the evolution of the dipole amplitude in which both linear radiative processes and non-linear recombination effects are systematically incorporated. The running-coupling BK (rcBK) equation  has the following simple form:
\begin{equation}
  \frac{\partial N_{F}(r,x)}{\partial\ln(x_0/x)}=\int d^2{\vec r_1}\
  K^{{\rm run}}({\vec r},{\vec r_1},{\vec r_2})
  \left[N_{F}(r_1,x)+N_{F}(r_2,x)
-N_{F}(r,x)-N_{F}(r_1,x) N_{F}(r_2,x)\right]\,,
\label{bk1}
\end{equation}
where $\vec r_2=\vec r- \vec r_1$. The evolution kernel $K^{{\rm run}}$ is given by Balitsky's
prescription \cite{bb}  with the running coupling. The explicit form of  $K^{{\rm run}}$ with details can be found in Refs.\,\cite{bb,rcbk}. The only external input for the rcBK non-linear equation is the initial condition for the evolution which is taken to have the following form motivated by McLerran-Venugopalan (MV) model \cite{mv},  
  \begin{equation}
\mathcal{N}(r,Y\!=\!0)=
1-\exp\left[-\frac{\left(r^2\,Q_{0s}^2\right)^{\gamma}}{4}\,
  \ln\left(\frac{1}{\Lambda\,r}+e\right)\right]\ ,
\label{mv}
\end{equation}
where $\Lambda=0.241$ GeV  \cite{j1,jav1}. The initial saturation scale $Q_{0s}$ (with $s=p,A$ for a proton and nuclear target)  at starting point of evolution (at $x_0=0.01$) and the parameter $\gamma$, are free parameters which are determined from a fit to other experimental measurements at small-x.  It was shown that inclusive single hadron data in pp collisions at RHIC  can be described with a initial saturation scale within $Q_{0p}^2= 0.168\div 0.336~\text{GeV}^2$ \cite{j-c,me-jamal1,raj}.  However, HERA data on proton structure functions prefers the lower value for the proton initial saturation scale $Q_{0p}^2\approx 0.168~\text{GeV}^2$ \cite{jav1}. One have also freedom to run $\gamma$ as a free parameter in the $\chi^2$ minimization and obtain its preferred value in a fit to HERA data. In order to investigate the uncertainties due to initial condition of the rcBK equation, we will consider the following three parameter sets which all provide an excellent fit to the HERA data for proton targets  \cite{j1,jav1}:
\begin{eqnarray} \label{set}
\text{set I}:&&\hspace{0.5cm} Q_{0p}^2=0.2\,\text{GeV}^2\hspace{0.80cm}\gamma=1 ,\nonumber\\
\text{set II}:&&\hspace{0.5cm} Q_{0p}^2=0.168\,\text{GeV}^2\hspace{0.5cm}\gamma=1.119,\nonumber\\
\text{set III}:&&\hspace{0.5cm} Q_{0p}^2=0.157\,\text{GeV}^2\hspace{0.5cm}\gamma=1.101.\
\end{eqnarray}
 In the MV model \cite{mv}, the parameter $\gamma$ in \eq{mv}  is $\gamma=1$. However, it has been recently shown \cite{dp}  that the effective value of $\gamma$ can be larger than one when the sub-leading corrections to the MV model are included~\cite{cubic-quartic}. 
The parameter $\gamma$ appears to be  also important in order to correctly reproduce the single inclusive particle spectra, and a larger value $\gamma>1$ is apparently preferable at large-$k_t$ \cite{j1,me-jamal1}.  

In our approach the difference between proton and nuclei originates from different initial saturation 
scales $Q_{0s}$ in the rcBK equation via Eq.\,(\ref{mv}).
In the case of inclusive hadron production in proton-nucleus collisions, due to theoretical uncertainties and rather large errors of the experimental data, 
it is not possible to uniquely fix the initial value of $Q_{0A}$. 
In the case of minimum-bias dAu collisions, the
initial nuclear (gold) saturation scale within $Q_{0A}^2=3 \div 4~Q_{0p}^2$ is consistent with the RHIC inclusive hadron production data~\cite{for-sup, j-c, me-jamal1,raj}. 
The extracted value of  $Q_{0A}$ is also consistent with the DIS data for nuclear targets \cite{jav1,raj}.  Here, we will also consider the uncertainties due to the initial condition of the rcBK equation for a nuclear target.  Note that $Q_{0A}$ should be considered as an impact-parameter averaged value since it was extracted from the minimum-bias data.  For the minimum-bias collisions, one may assume that the initial saturation scale of a nuclei with atomic mass number A, scales linearly with $A^{1/3}$, namely we have $Q_{0A}^2=cA^{1/3}~Q_{0p}^2$ where the parameter $c$ is fixed from a fit to data. In Ref.\,\cite{raj}, it was shown that NMC data can be described with $c\approx 0.5$.  

We will use the NLO MSTW 2008 PDFs \cite{mstw} and the 
NLO KKP FFs \cite{kkp}. For the photon fragmentation function, we will use the full leading log parametrization \cite{own,ffp}.  We assume the factorization scale $Q$ in the FFs and the PDFs to be equal and its value is taken to be  $q_t$ and $k_t$ for the semi-inclusive and inclusive prompt photon production, respectively.

\subsection{Direct and fragmentation prompt photon in pp and pA collisions at RHIC and the LHC} 

We start by considering direct and fragmentation photon production in pA collisions at RHIC and the LHC. 
In nuclear collisions, nuclear effects on single particle production are usually evaluated in terms of ratios of particle yields in pA and pp collisions (scaled with a proper normalization), the so-called  nuclear modification factor $R_{pA}$. The nuclear modification factor $R_{p(d)A}$ is defined as
\begin{eqnarray}\label{rpa}
R_{dA}^{\gamma}&=&\frac{1}{2\, N_{coll}}\frac{dN^{d A \rightarrow \gamma X}}{d^2p_T d\eta}/
\frac{dN^{p p \rightarrow \gamma X}}{d^2p_T d\eta}, \nonumber\\ 
R_{pA}^{\gamma}&=&\frac{1}{N_{coll}}\frac{dN^{p A \rightarrow \gamma X}}{d^2p_T d\eta}/
\frac{dN^{p p \rightarrow \gamma X}}{d^2p_T d\eta},\
\end{eqnarray}
where  the yield  $\frac{dN^{p(d) A(p) \rightarrow \gamma X }}{d^2p_T d\eta}$ can be calculated from the invariant cross-section given in \eq{pho4}. The normalization constant $N_{coll}$ is the number of binary proton-nucleus collisions. We take $N_{coll}=3.6, 6.5$ and $7.4$ at $\sqrt{s}=0.2, 4.4$ and $8.8$ TeV, respectively \cite{ncoll}. In order to compare our predictions for $R_{pA}^{\gamma}$ with the experimental value, one should take into account possible discrepancy between our assumed normalization $N_{coll}$ and the experimentally measured value for $N_{coll}$ by rescaling our curves. Again we expect that some of the theoretical uncertainties, such as sensitivity to $K$ factors (which effectively incorporates the missing higher order corrections), will drop out in $R_{p(d)A}^{\gamma}$. 

\begin{figure}[t]                                       
\includegraphics[width=8.5 cm] {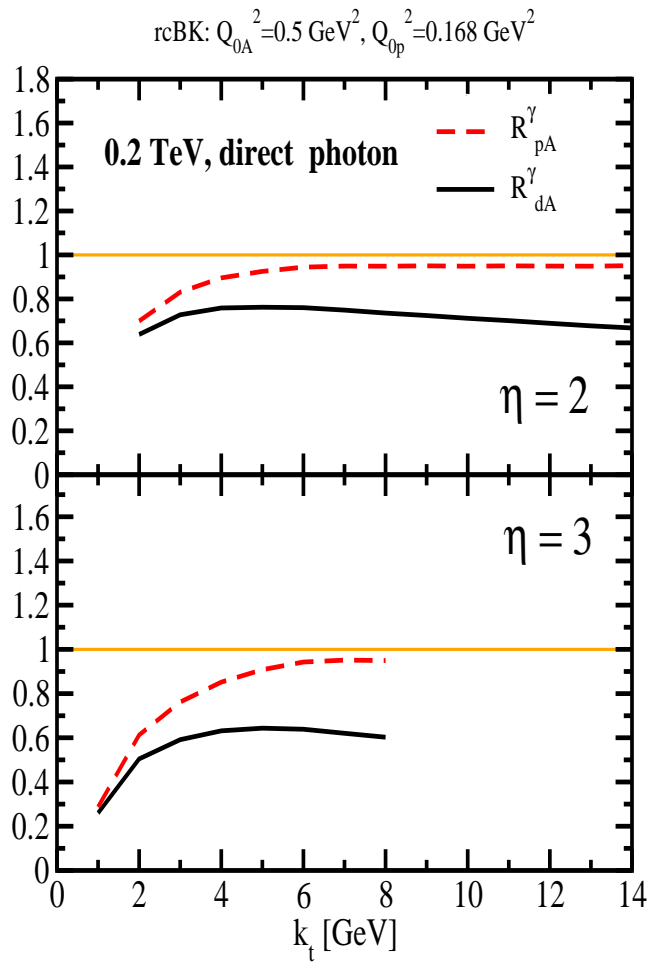}                                   
\caption{Nuclear modification factor for direct photon production in minimum-bias pA (dashed line) and dA (solid line) collisions at RHIC ($\sqrt{S}=0.2$ TeV) at $\eta = 2, 3$. The curves are obtained from \eq{pho4} using the solution to rcBK equation with the initial saturation scale $Q_{0p}^2=0.168~\text{GeV}^2$ for proton and $Q_{0A}^2=3Q_{0p}^2$  for a nucleus (gold). }
\label{fig-p-vs-d}
\end{figure}

In \fig{fig-p-vs-d} we show the nuclear modification factor for both pA and dA collisions at RHIC. This is to facilitate a comparison of and to distinguish between the genuine saturation effects in the target nucleus from isospin effects in the projectile
deuteron. 
Clearly there is a large difference between a proton and a deuteron projectile as far as prompt photon production is concerned. This difference is more pronounced in the forward rapidity region and at high transverse momentum where one probes the quark content of the projectile. This is due to difference between the up and down quark distributions of a proton (note that nuclear effects in the wave function of a deuteron are ignored as they are known to be small). This is a well known effect, nevertheless, for sake of clarity and to illustrate the difference between inclusive hadron production and QED probe namely prompt photon production, we illustrate this in some detail. In case of photon production, the production cross section is weighed by the the charged squared of a given quark flavor. For example (assuming only two flavor), for a proton projectile this is given by $e_q^2\, f_{q/p} = (2/3)^2 \, u_p + (1/3)^2\, d_p$ where $u_p, d_p$ denote the distribution functions of up and down quarks in a proton. Ignoring nuclear effects in a deuteron, we assume a deuteron is a system of free proton and a neutron in which case the corresponding expression is $e_q^2\, f_{q/d} = (2/3)^2 \, u_p + (1/3)^2\, d_p + (2/3)^2\, u_n + (1/3)^2\, d_n$ where $u_n, d_n$ denote the distribution functions of up and down quarks in a neutron. Assuming isospin symmetry gives $u_n = d_p$ and $d_n = u_p$ which leads to $(5/9)\, [u_p + d_p]$ for a deuteron. Comparing this expression with two times that of a proton, the relative contribution of up quarks in a deuteron ($5/9$) is smaller than that of a twice a proton ($8/9$). Since there are more up quarks than down quarks (by a factor of $2\div 3$ in this kinematics) in a proton, and their relative weight is smaller, this leads to a further reduction of $R_{dA}$ as compared with $R_{pA}$ in prompt photon production, see \fig{fig-p-vs-d}. We note that in the absence of the charged squared factor, which is the case for inclusive hadron production, one would get $d = p + n = 2\, p$ as one should since possible nuclear effects in a deuteron are ignored here.  At the LHC the isospin effect is absent since the same projectile is used for the pp and pA collisions. This helps to understand the physics of QCD saturation more clearly, as the suppression of the signal will not be contaminated with isospin effect.

\begin{figure}[t]                                       
                                  \includegraphics[width=8.5 cm] {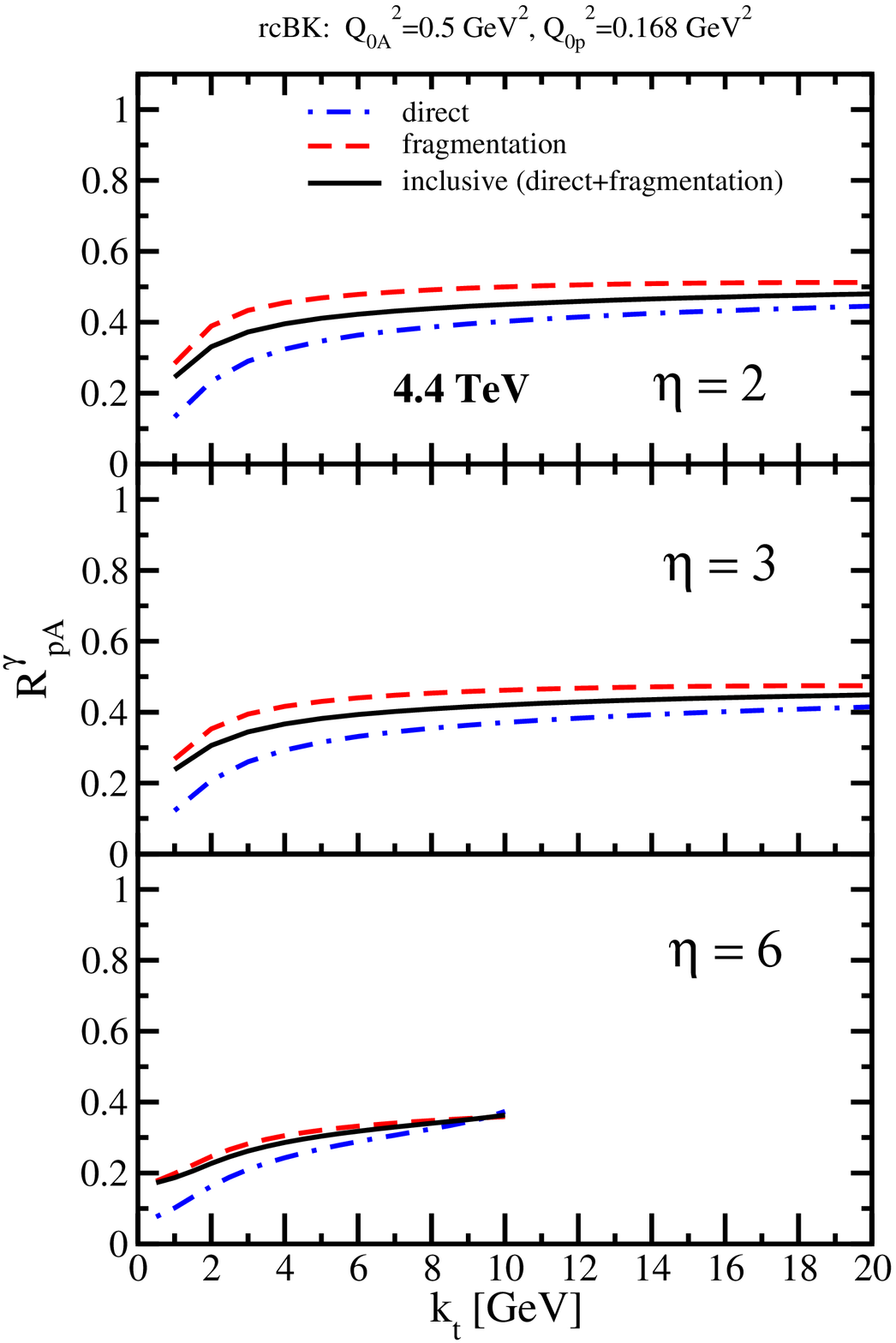} 
                                  \includegraphics[width=8.5 cm] {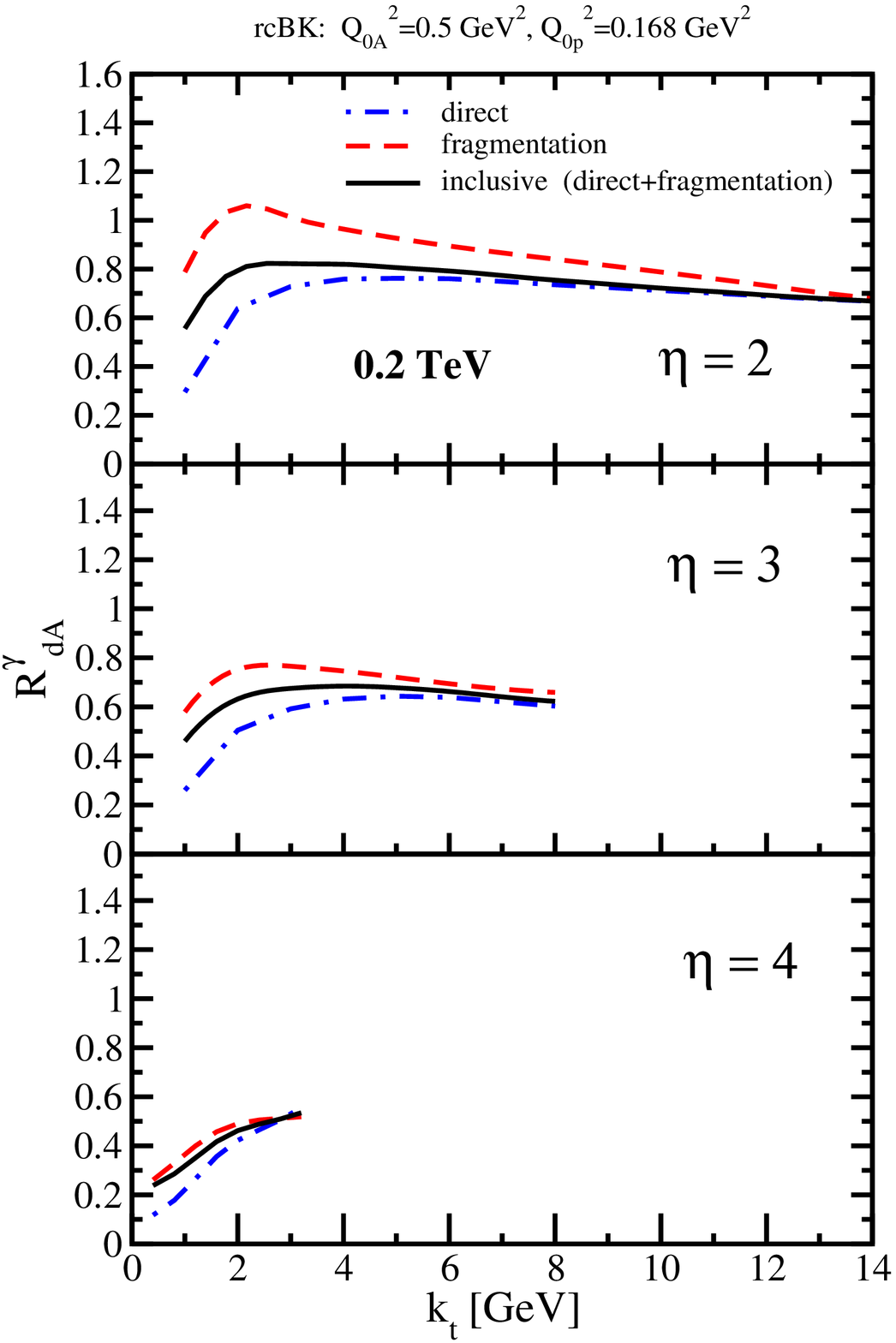}                                  
\caption{Nuclear modification factor for direct, fragmentation and inclusive prompt photon production in minimum-bias p(d)A collisions at RHIC $\sqrt{S}=0.2$ TeV (right) and the LHC $\sqrt{S}=4.4$ TeV (left) energy at various rapidities. The curves are the results obtained from \eq{pho4} and the solution to rcBK equation with  the initial saturation scale $Q_{0p}^2=0.168~\text{GeV}^2$ for a proton and $Q_{0A}^2=3Q_{0p}^2$  for a nucleus  (gold), corresponding to set II in \eq{set}. }
\label{fig-r1}
\end{figure}

\begin{figure}[t]                            
                                 \includegraphics[width=8.5 cm] {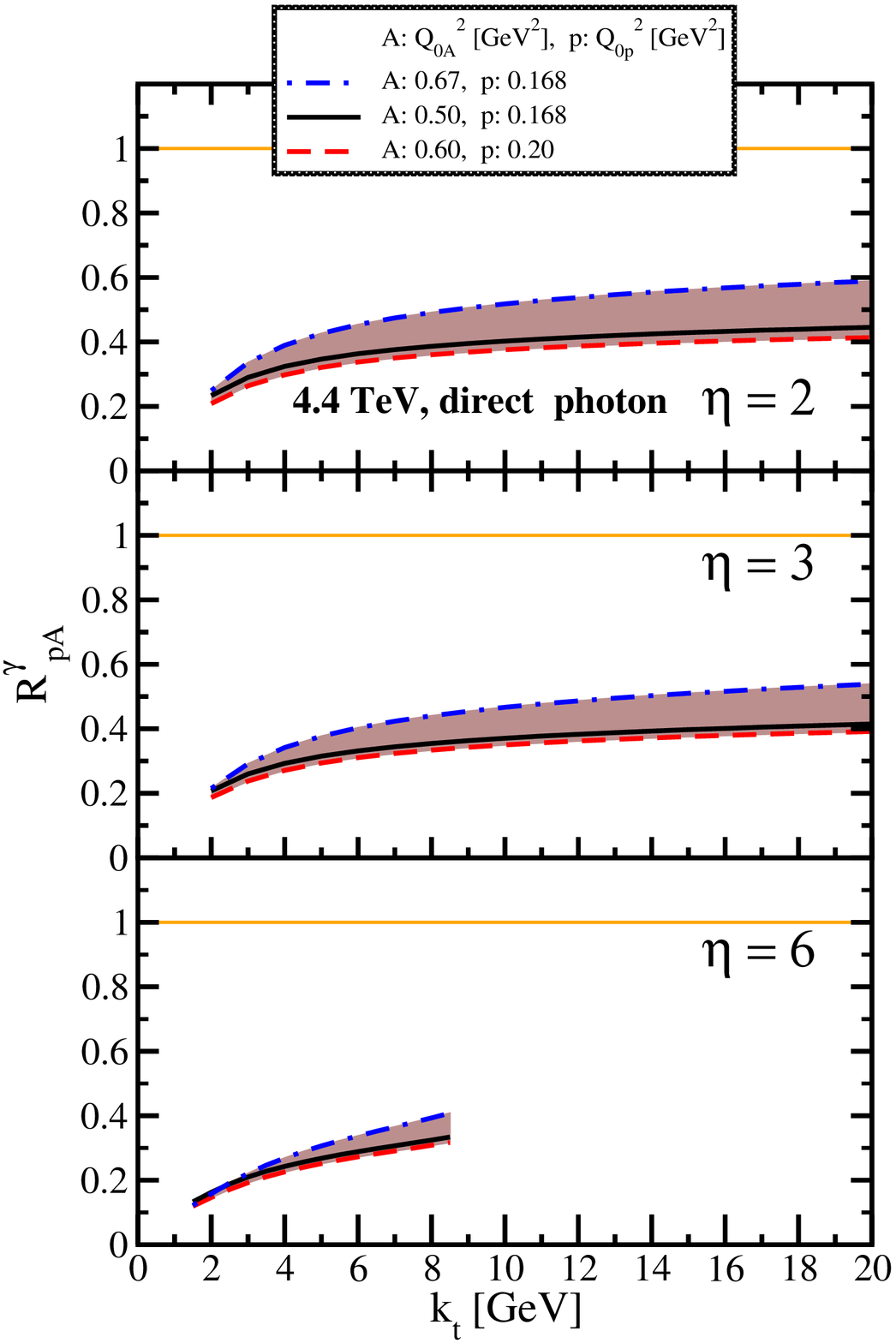} 
                                  \includegraphics[width=8.5 cm] {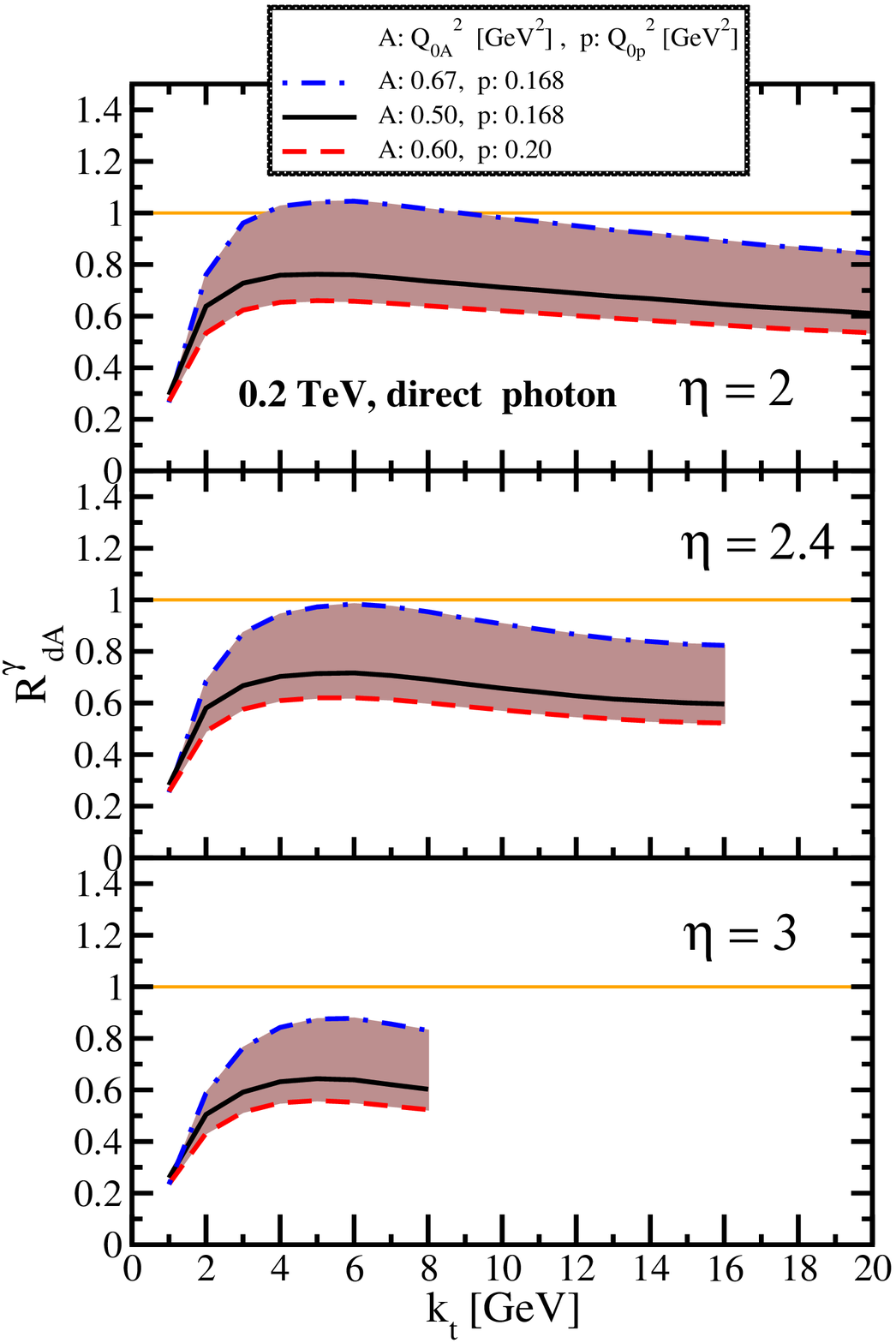}                                  
\caption{Nuclear modification factor for direct photon production in p(d)A collisions at various rapidities at RHIC $\sqrt{S}=0.2$ TeV (right) and the LHC $\sqrt{S}=4.4$ TeV energy (left). The curves are the results obtained from \eq{pho4} and the solution to rcBK equation using different initial saturation scales for a proton $Q_{0p}$ and a nucleus $Q_{0A}$. The band shows our theoretical uncertainties arising from allowing a variation of the initial saturation scale of the nucleus in a range consistent with previous studies of DIS structure functions as well as particle 
production in minimum-bias pp, pA and AA collisions in the CGC formalism, see the text for the details.}
\label{fig-r2}
\end{figure}

\begin{figure}[t]                            
                                \includegraphics[width=8.5 cm] {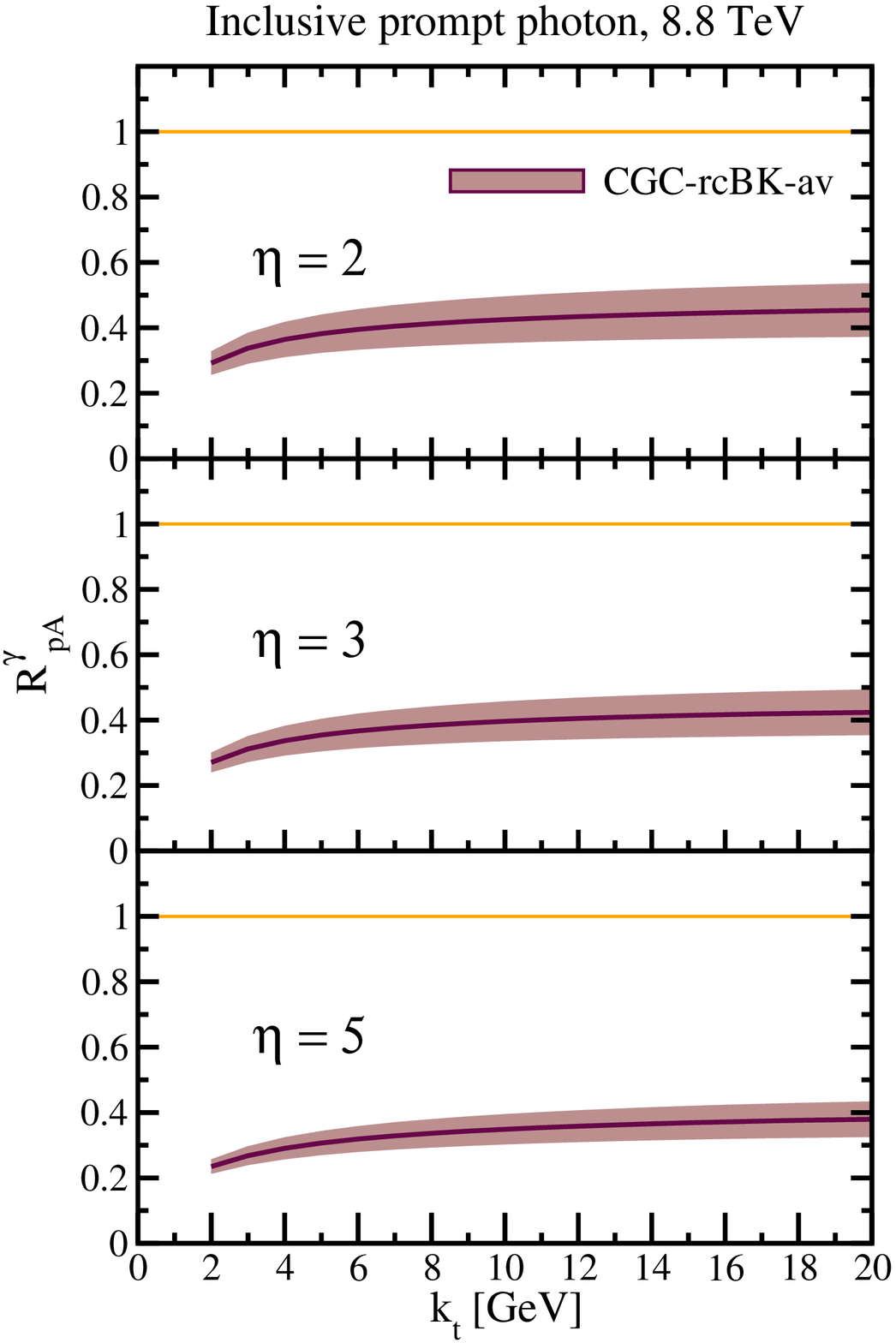}                                  
                                 \includegraphics[width=8.5 cm] {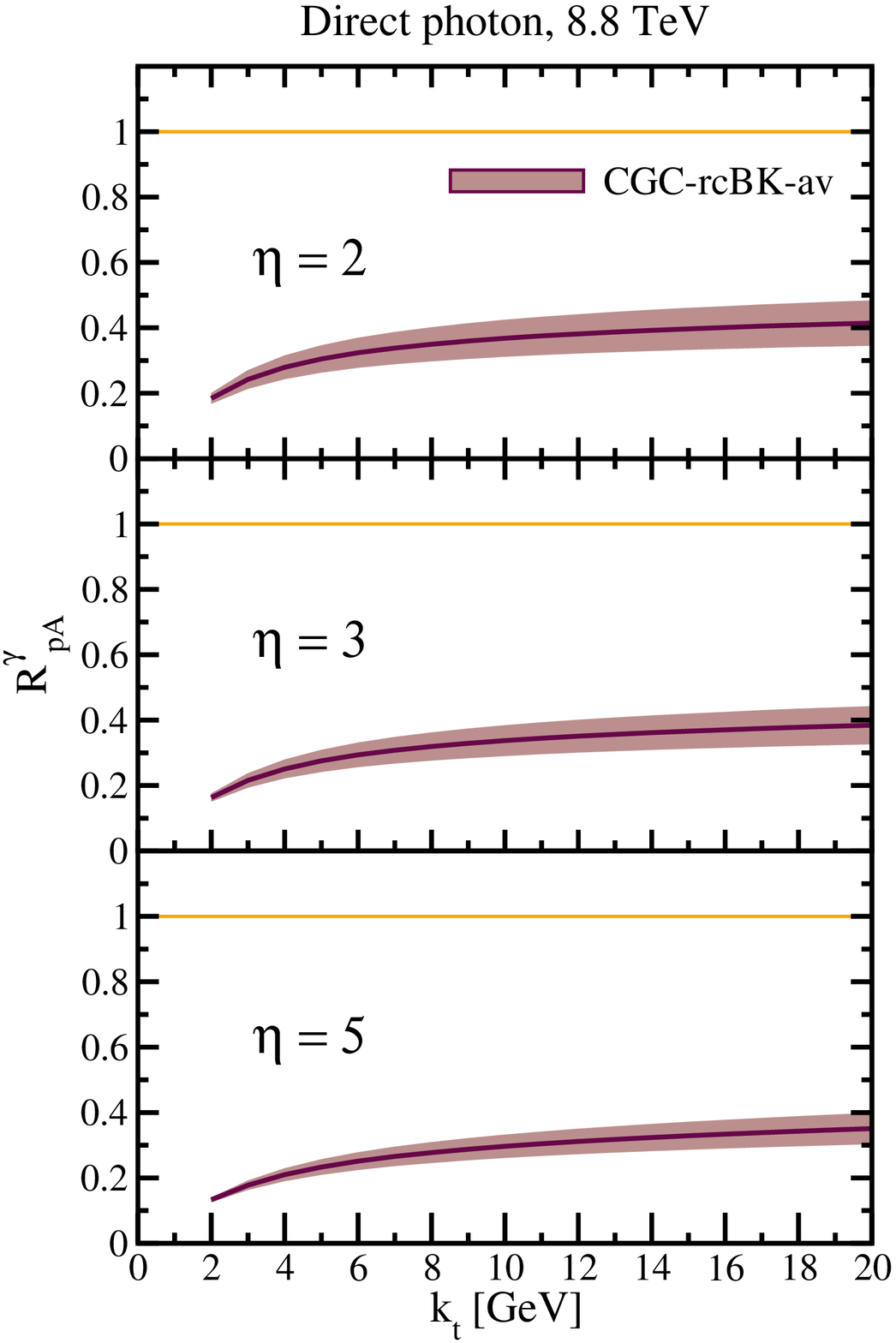}                                  
\caption{Nuclear modification factor for direct photon (right) and inclusive prompt photon (left) production in pA collisions at various rapidities a the LHC $\sqrt{S}=8.8$ TeV energy. The band (CGC-rcBK-av) similar to \fig{fig-r2} corresponds to the results obtained from \eq{pho4} and the solutions to the rcBK evolution equation using different initial saturation scales for a proton $Q_{0p}$ and a nucleus $Q_{0A}$,  see the text for the details.}
\label{fig-r22}
\end{figure}

\begin{figure}[t]     
                      \includegraphics[width=8.5 cm] {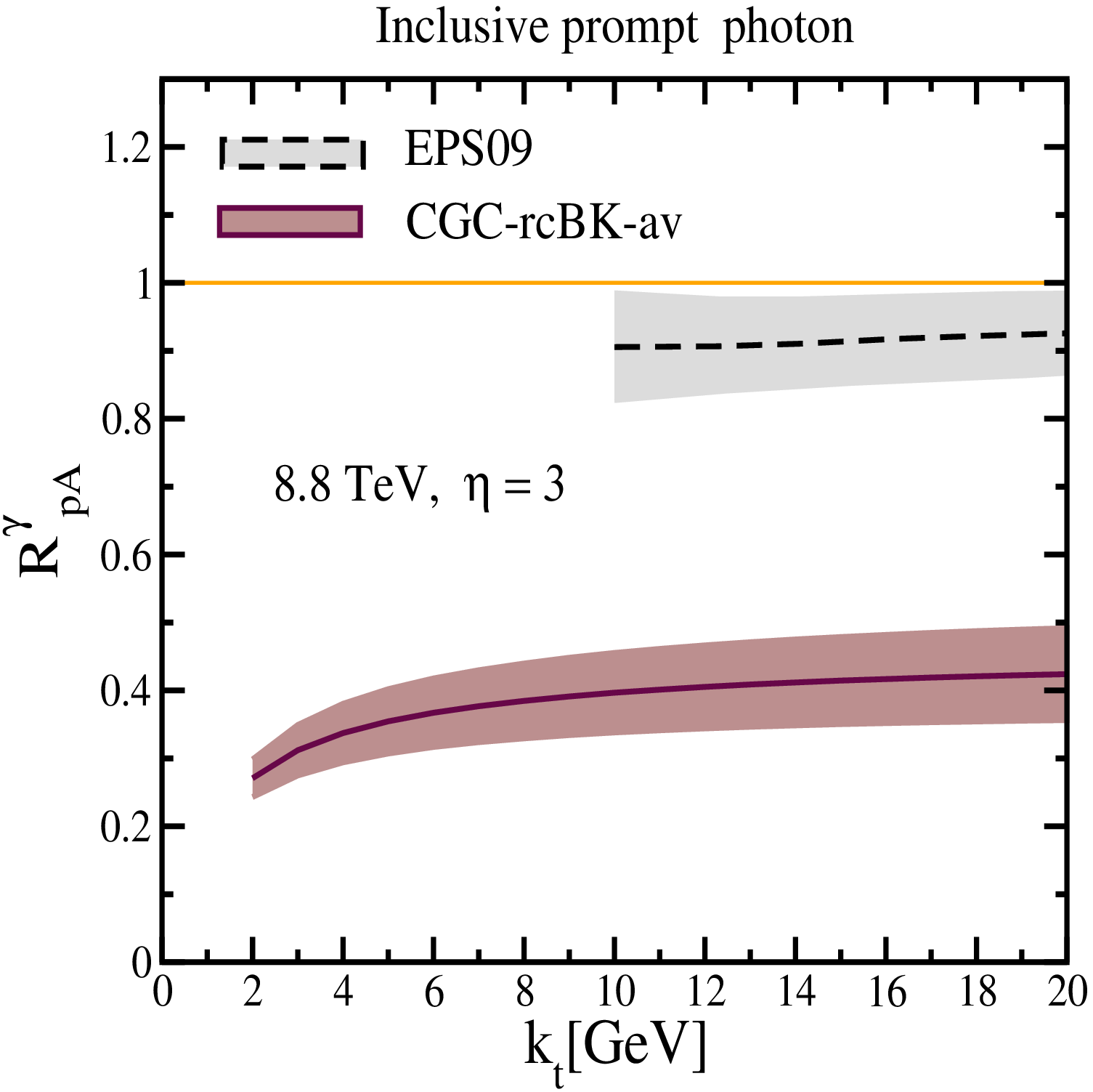}                            
                                  \includegraphics[width=8.5 cm] {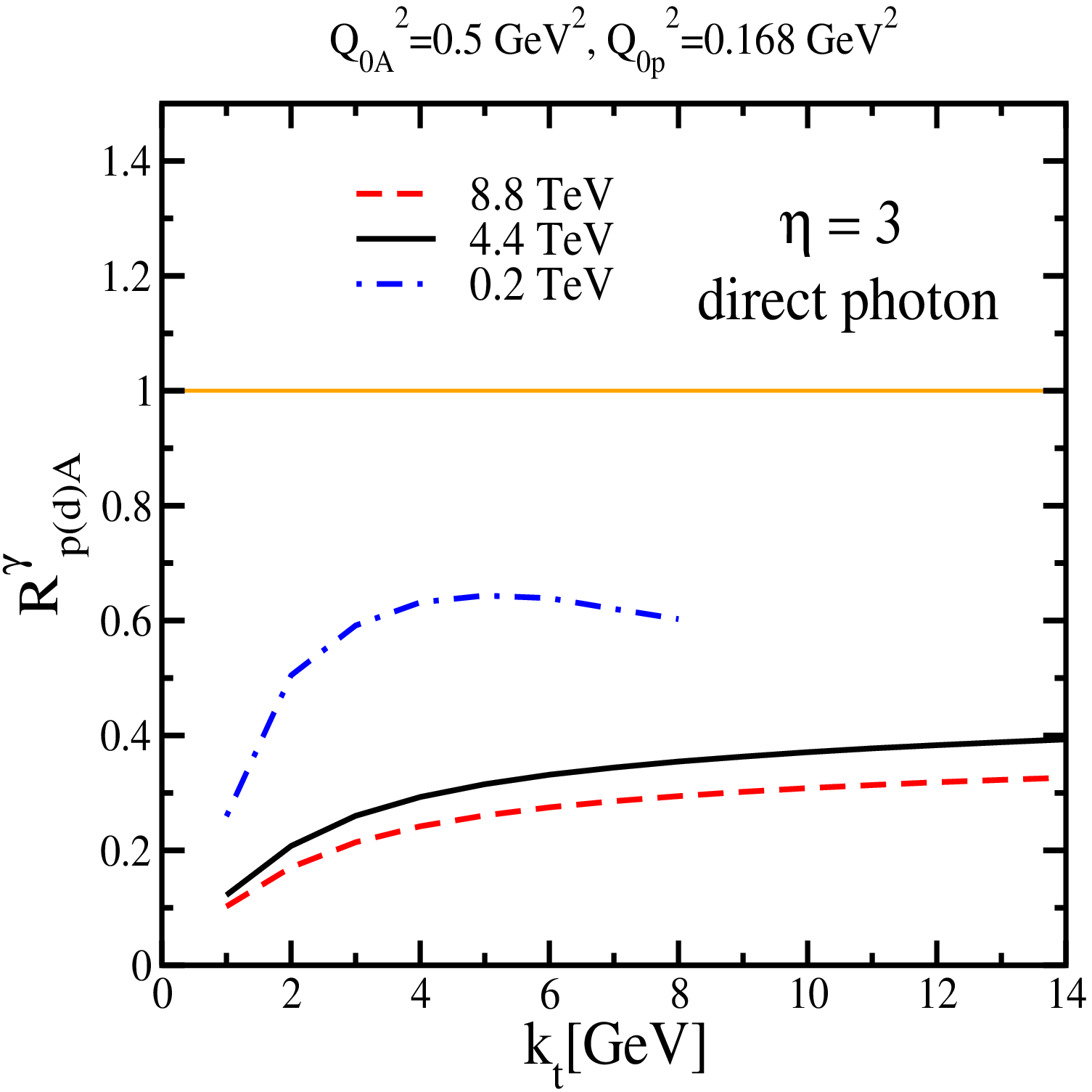}                                                  
\caption{Right: Nuclear modification factor for direct photon production  at $\eta=3$ in minimum-bias dA  $\sqrt{S}=0.2$ TeV (RHIC)  and pA $\sqrt{S}=4.4, 8.8$ TeV (LHC) collisions.  The curves are the results obtained from \eq{pho4} and the solution to rcBK equation with the initial saturation scale $Q_{0p}^2=0.168~\text{GeV}^2$ for a proton and $Q_{0A}^2=3Q_{0p}^2$  for a nucleus. Left: Comparison of the inclusive prompt photon nuclear modification factor predictions from the CGC (in this paper) and the standard collinear factorization approach \cite{pqcd-models}. The band CGC-rcBK-av is the same as in \fig{fig-r22}.   }
\label{fig-r3}
\end{figure}

In \fig{fig-r1}, we show the minimum-bias nuclear modification factor for the direct, fragmentation and the inclusive prompt photon production at RHIC and the LHC energies $\sqrt{S}=0.2, 4.4$ TeV at various rapidities $\eta$ obtained from Eqs.\,(\ref{pho4}, \ref{rpa}) supplemented  with rcBK solution \eq{bk1} with the initial saturation scale for proton $Q_{0p}^2\approx 0.168~\text{GeV}^2$ and nuclei $Q_{0A}^2=3 Q_{0p}^2$.  It is seen that the nuclear modification $R_{p(d)A}^{\gamma}$ for the fragmentation photon is bigger than the direct and inclusive prompt photon. This is what we expected in our picture since direct photon cross-section in \eq{pho2} probes the target structure function at lower transverse momentum $k_t$ (and consequently lower x) than the fragmentation part with transverse momentum $k_t/z$ and therefore is more sensitive to the suppression of structure function and the saturation effect. 
However, as we increase the energy the enhancement of the fragmentation photon $R_{p(d)A}^{\gamma}$ at RHIC will be also replaced with suppression at the LHC, see  \fig{fig-r1} top panel. This is simply due to the fact that both the fragmentation and the direct part \eq{pho2}  depend on the color dipole forward amplitude which encodes the small-x dynamics and at higher energy, the small-x evolution leads to suppression of $R_{p(d)A}^{\gamma}$.  

In a collider experiment such as the LHC, the secondary photons coming from the decays of hadrons, overwhelm  the inclusive prompt photon measurements with order of magnitudes. In order to reject the background, isolation cuts are imposed \cite{facp}. Contribution of fragmentation prompt photon is reduced by imposing an isolation cut\footnote{If we assume that $p_c$ is the total transverse momentum of a fragmentation jet, the photon's energy is then $E_\gamma=z p_c$ and the total hadronic energy within the jet is  $E_h=(1-z)p_c$. By isolation cut criterion, the hadronic energy does not have to be more than $\epsilon E_\gamma$ in the isolation cone. This gives the lower limit of $z$ (or $x_q$ convolution) in \eq{pho5} integral, namely $z_c=1/(1+\epsilon)<z$. Given that the integrand of fragmented part is proportional to $1/z$ and dominated at lower limit of integrand, we expect that the isolation cut reduces the fragmentation contribution more severely than the direct one.}. A proper incorporation of the isolation cut criterion in our framework is beyond the scope of this paper. However, from 
 \fig{fig-r1} it is seen that at higher energy at forward collisions, $R_{p(d)A}^{\gamma}$ for direct and single inclusive prompt photon becomes remarkably similar, indicating that to a good approximation, one may assume that the nuclear modification factor for direct and isolated prompt photon are equal.

In \fig{fig-r2}, we show the minimum-bias nuclear modification factor for the direct photon production at RHIC and the LHC energies $\sqrt{S}=0.2, 4.4$ TeV at various rapidities $\eta$ obtained from rcBK solutions \eq{bk1} with the initial proton saturation scale $Q_{0p}^2\approx 0.168$ and $0.2\,\text{GeV}^2$ corresponding to parameter sets I and II in \eq{set}.
For nuclear target in minimum-bias collisions, we take two initial saturation scales for nuclei (gold and lead) $Q_{0A}^2=3\div4 Q_{0p}^2$ which are extracted from a fit to other experimental data on heavy nuclear target \cite{j-c, me-jamal1,raj}. For a proton target, we have checked that parameter sets II and III give similar results for $R_{p(d)A}^{\gamma}$ with better than $10\% $ accuracy. Therefore, in  \fig{fig-r2} we only show results obtained from two parameter sets I and II in \eq{set}. The band in \fig{fig-r2} 
shows our uncertainties arising from a variation of the initial saturation scale of 
the nucleus in a range consistent with previous studies of DIS structure functions as well as particle 
production in minimum-bias pp, pA and AA collisions in the CGC formalism. One may therefore expect that the possible effects of fluctuations (of nucleons in a nucleus) on particle production is effectively contained in our error band.

From \fig{fig-r2}, it is seen that the nuclear modification for direct photon production is very sensitive to the initial saturation scale in proton and nuclei.  However, this uncertainties will be reduced for more forward collisions at higher energy at the LHC. The same effect has been observed for the inclusive hadron production in pA collisions \cite{me-jamal1}. This clearly indicates that the nuclear modification in p(d)A collisions is a sensitive probe of saturation effects and $R_{p(d)A}^{\gamma}$ measurements for direct photon and inclusive hadron provide crucial complementary information about initial saturation scale and small-x evolution dynamics.  In \fig{fig-r2}, 
it is seen that at a fixed rapidity and energy for a fixed initial saturation scale for proton $Q_{0p}$, a bigger initial saturation scale for nuclei  $Q_{0A}$ leads to a bigger broadening and consequently  enhances the cross-section and $R_{p(d)A}^{\gamma}$  if  $N_{coll}$ is kept fixed.

 In \fig{fig-r22},  we show our predictions for $R_{pA}$ for direct photon (right) and inclusive prompt photon (left) production in pA collisions at various rapidities a the LHC $\sqrt{S}=8.8$ TeV energy. The band (CGC-rcBK-av) similar to \fig{fig-r2} corresponds to the results obtained from \eq{pho4} with the solutions of the rcBK evolution equation (\ref{bk1}).

In \fig{fig-r3} (right), we compare $R_{p(d)A}^{\gamma}$ for direct photon at $\eta=3$ for RHIC energy $\sqrt{S}=0.2$ TeV and the LHC energies $4.4, 8.8$ TeV. It is seen that the suppression of $R_{pA}^{\gamma}$ at the LHC is larger compared to $R_{dA}$ at RHIC and persists at higher transverse momentum. This larger suppression is even more impressive given that fact that a good amount of the observed suppression of $R_{dA}$ at RHIC is due to the projectile being a deuteron rather than a proton. In \fig{fig-r3} (left), we compare the CGC prediction (CGC-rcBK-av)  obtained here with the collinear factorization result (EPS09) \cite{pqcd-models} for inclusive prompt photon $R_{pA}^{\gamma}$  at $\eta=3$ at the LHC. It is seen that the LHC measurements of the inclusive prompt photon at forward rapidities can discriminate between the collinear (standard parton model) and the CGC approach. 

Some words of caution are in order here.  Strictly speaking our formalism is less reliable for collisions at around mid-rapidities and high transverse momenta. This is due to the fact that our formula is valid for asymmetric collisions like pA or pp collisions at forward rapidities when a projectile can be treated in the standard collinear approximation while for the target we systematically incorporated the small-x re-summation (at the leading log approximation) effects.  Note however, for the case of pp collisions (our reference for $R_{dA}^{\gamma}$) at RHIC, the saturation scale of target proton is rather small, and it is not clear that the CGC formulation will be applicable. Moreover, our parameter sets in \eq{set} was obtained from a fit to HERA data at small-x  $x<0.01$ and for virtualities $Q^2\in[0.25,40] \,\text{GeV}^2$ \cite{jav1}. Therefore, our predictions are less reliable at high-$k_t$ ($k_t> 6\div 7$ GeV).

\begin{figure}[t]       
                                \includegraphics[width=8.1 cm] {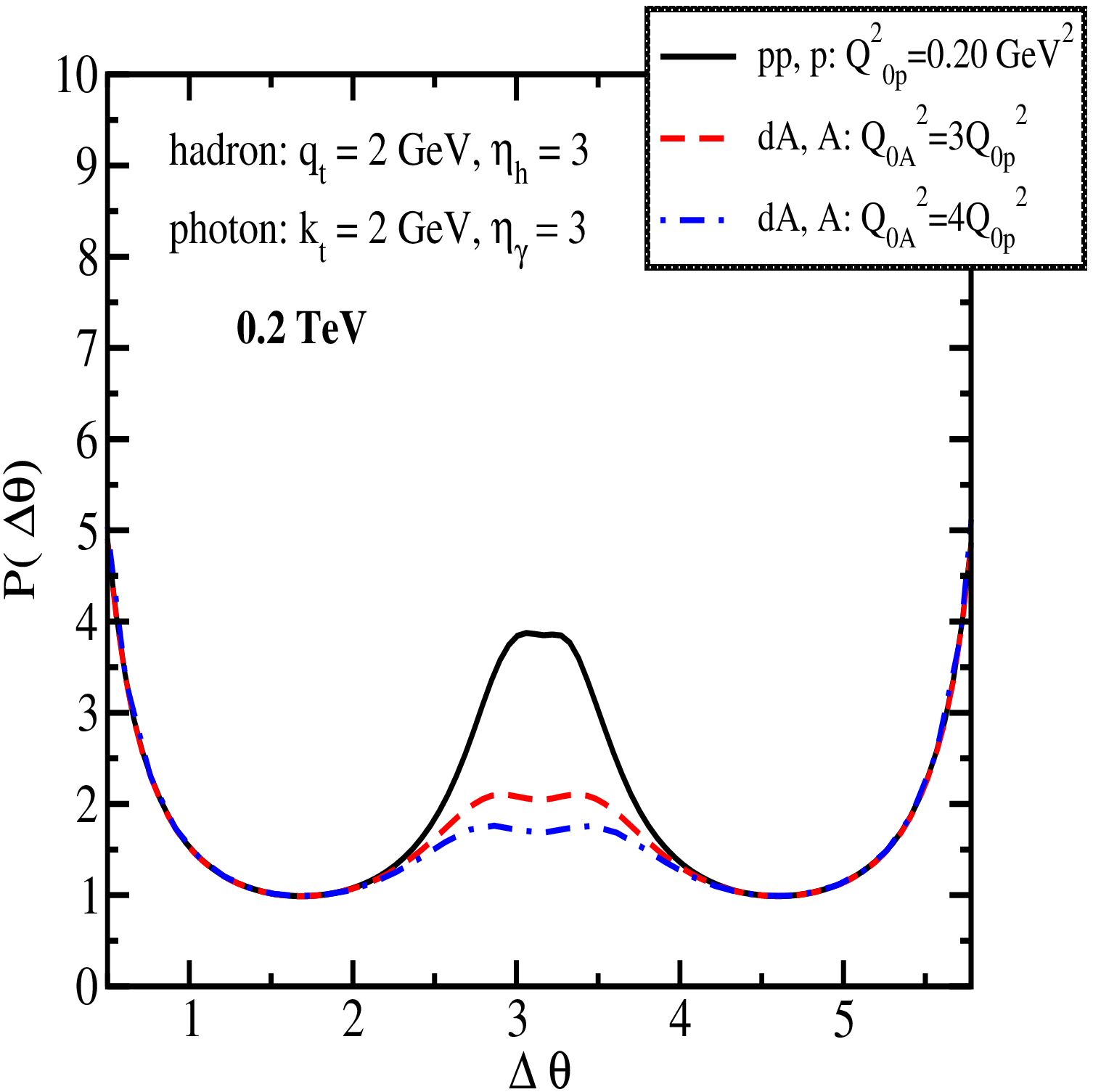} 
                                \includegraphics[width=8.1 cm] {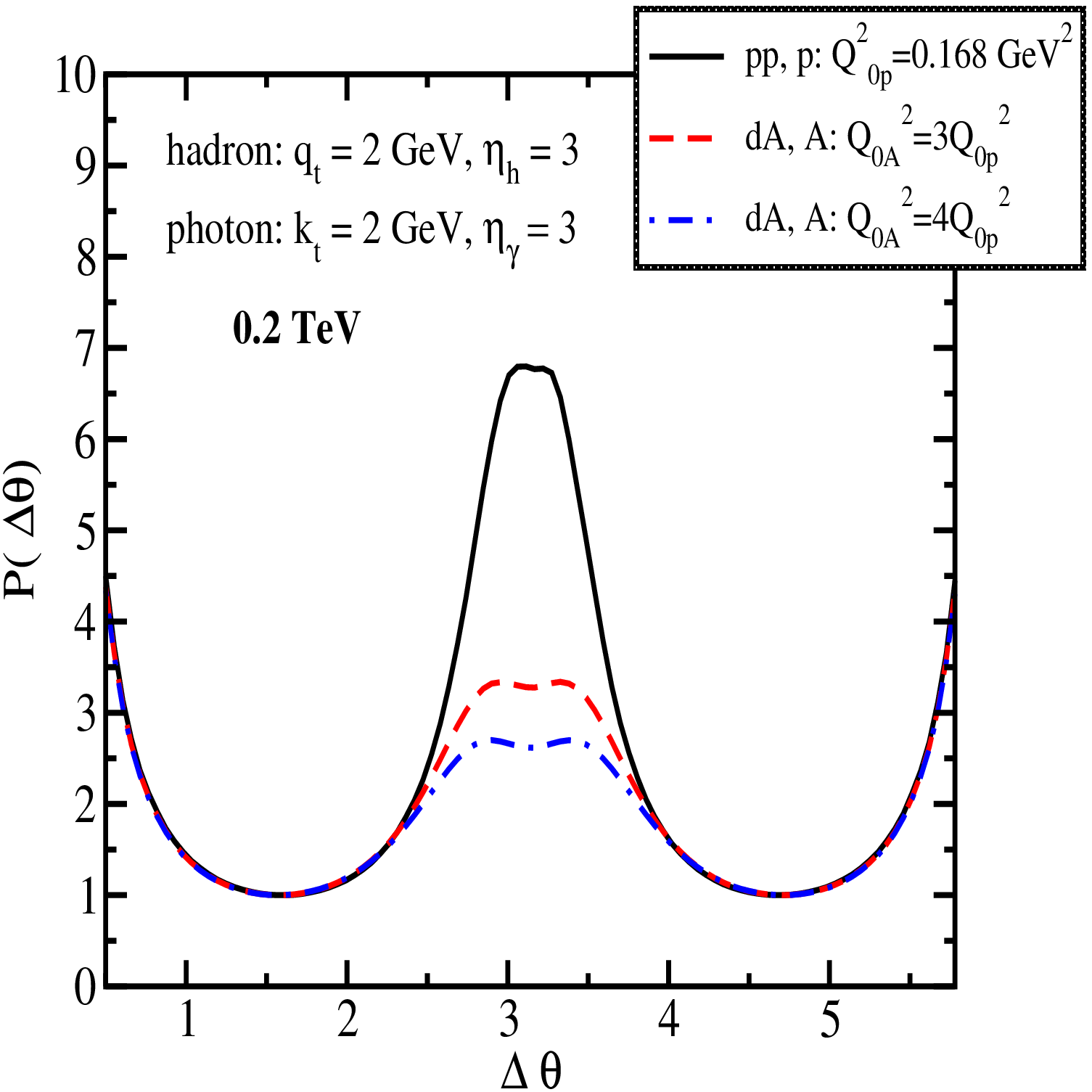}                 
                                  \includegraphics[width=8.1 cm] {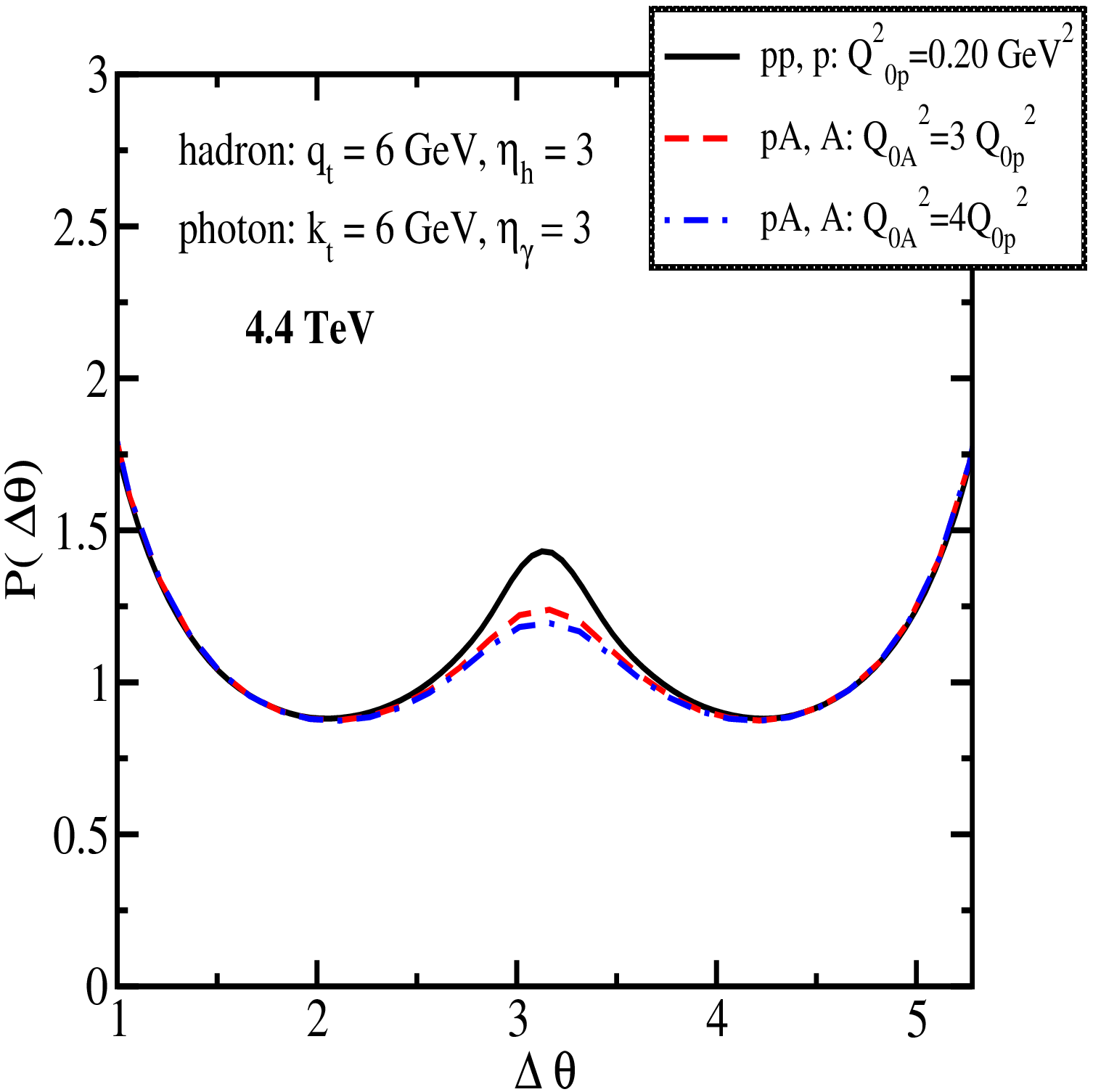} 
                                  \includegraphics[width=8.1 cm] {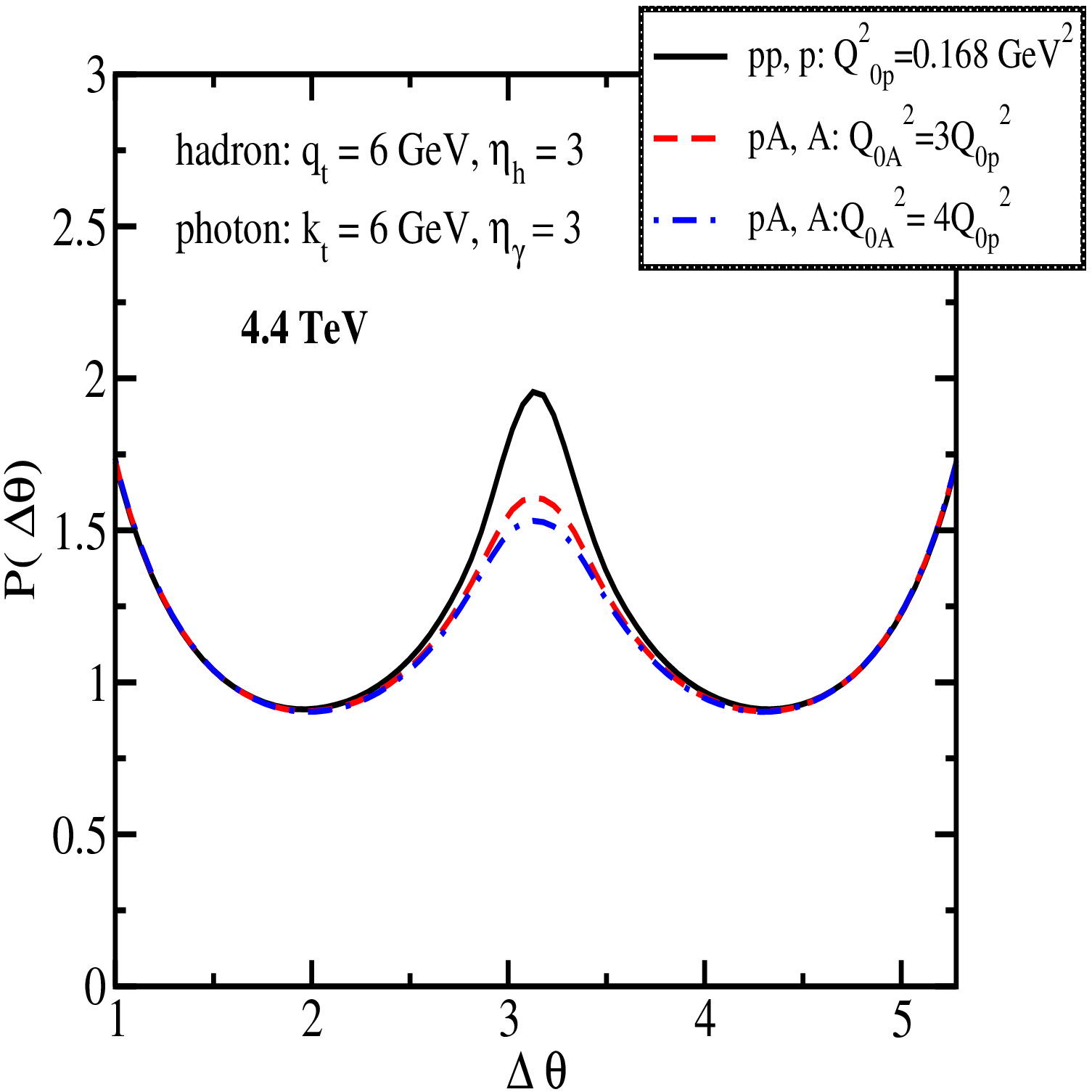}  
\caption{The relative azimuthal correlation $P(\Delta \theta)$ defined in \eq{az} for minimum-bias p(d)A and pp collisions at RHIC $\sqrt{S}=0.2$ TeV (upper) and the LHC $\sqrt{S}=4.4$ TeV energy (lower) obtained from the rcBK solutions with different initial saturation scales. }
\label{fig1-1}
\end{figure}

\subsection{Prompt photon-hadron correlations at RHIC and the LHC; the signature of saturation} 

We now focus on azimuthal angle $\Delta \theta$ correlations of the prompt photon-hadron spectrum, where the 
angle $\Delta \theta$ is the difference between the azimuthal angle of the measured hadron and single prompt photon. 
We present our predictions for semi-inclusive prompt photon-hadron (for hadron we consider only neutral pion here) production at RHIC and the LHC in pp and p(d)A collisions in terms of  $P(\Delta \theta)$ defined as follows, 
\begin{equation}\label{az}
P(\Delta \theta)={d\sigma^{p(d)\, T \rightarrow h(q)\,\gamma(k)\, X}
\over d^2\vec{b_t}\, dk_t^2\, dq_t^2\, dy_{\gamma}\, dy_l\, d\theta} [\Delta \theta]/{d\sigma^{p(d)\, T \rightarrow h(q)\,\gamma(k)\, X}
\over d^2\vec{b_t}\, dk_t^2\, dq_t^2\, dy_{\gamma}\, dy_l\, d\theta} [\Delta \theta= \Delta \theta_c],
\end{equation}
where the prompt photon-hadron cross-section in above expression is given in \eq{qh-f}. This definition has a simple meaning of the probability of, the single semi-inclusive prompt photon-hadron production at a certain kinematics and angle  $\Delta \theta$ given the  production with the same kinematics at  a fixed reference angle $\Delta \theta_c$. We take  $\Delta \theta_c=\pi/2$.  As will show the 
$P(\Delta \theta)$ defined in this way has a non-trivial structure and can probe the physics of small-x and gluon saturation.
In principle, one is free to chose a different reference angle $\Delta \theta_c$, however any value $\Delta \theta_c<<\pi$ will only change the normalization rather than the main picture. The advantage of the above definition for the azimuthal correlations is that it is experimentally easier to measure as it does not require a different experimental setup and run for the trigger or reference.  Moreover, in dA collisions at RHIC, the isospin effect in $P(\Delta \theta)$ will drop out via normalization\footnote{We checked that numerically the isospin effect brings less than $2\%$ contribution to the azimuthal correlation defined via \eq{az}. Therefore, due to our particular definition of  $P(\Delta \theta)$ in \eq{az}, the differences between a deuteron and a proton projectile are negligible unlike the prompt photon production case.}  and this facilitates to single out the importance of the saturation effect at forward rapidities in contrast to the nuclear modification factor $R_{dA}^{\gamma}$.

\begin{figure}[t]                                                            
                                  \includegraphics[width=8.5 cm] {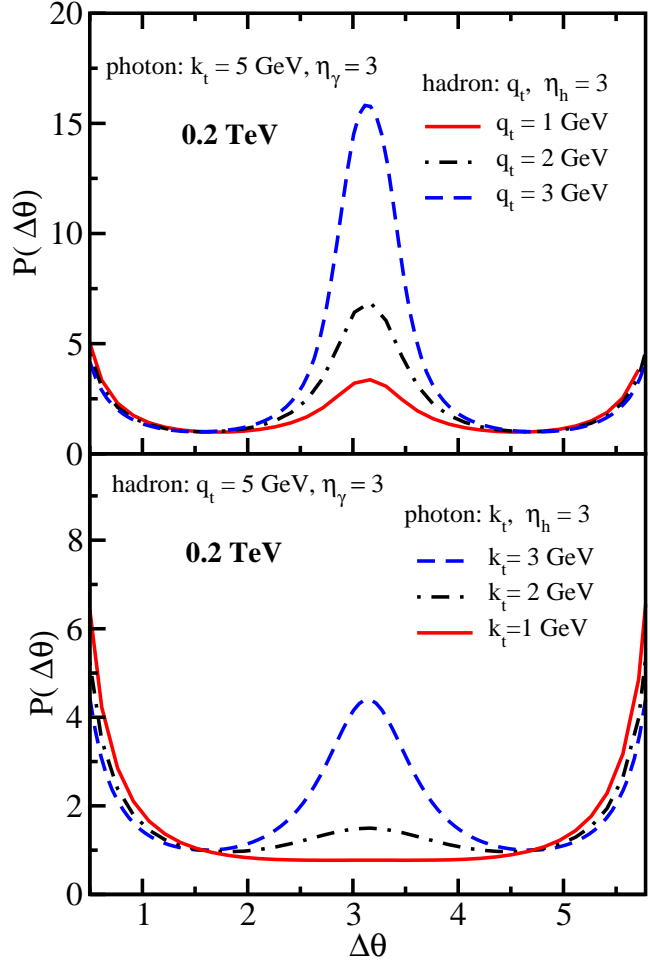}     
\caption{ The relative azimuthal correlation $P(\Delta \theta)$ defined in \eq{az} for minimum-bias dA collisions at RHIC $\sqrt{S}=0.2$ TeV  for two different windows of kinematics of transverse momentum of produced prompt photon $k_t$ and hadron (neutral pion) $q_t$ at fixed rapidity $\eta_h=\eta_{\gamma}=3$. The curves are the results obtained from the rcBK equation solution with  the initial saturation scale $Q_{0p}^2=0.168~\text{GeV}^2$ for proton and $Q_{0A}^2=3Q_{0p}^2$  for a nuclei.  }
\label{fig2}
\end{figure}

\begin{figure}[t]                                                            
                                  \includegraphics[width=8.1 cm] {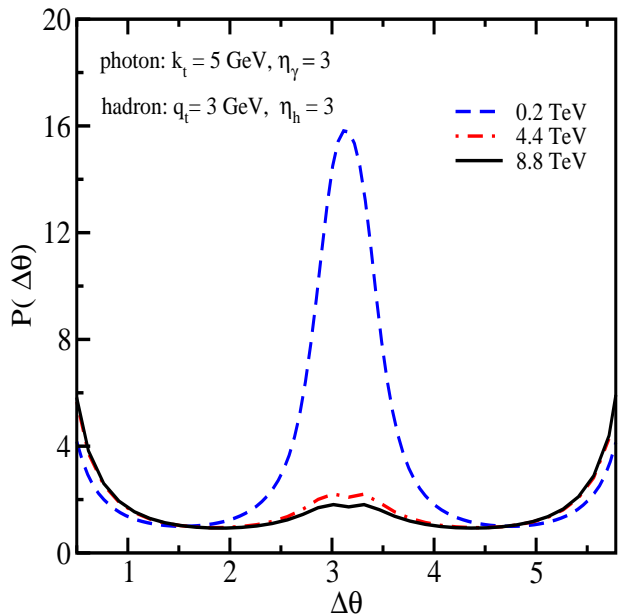}       
\caption{ The relative azimuthal correlation $P(\Delta \theta)$ for minimum-bias p(d)A  collisions at different energies $\sqrt{S}=0.2, 4.4, 8.8$ TeV for a fixed rapidity $\eta_h=\eta_{\gamma}=3$ and transverse momentum of the produced prompt photon $k_t=5$ GeV and hadron (neutral pion) $q_t=3$ GeV. The curves are the results obtained from the rcBK equation solution with  the initial saturation scale $Q_{0p}^2=0.168~\text{GeV}^2$ for proton and $Q_{0A}^2=3Q_{0p}^2$  for a nuclei}
\label{fig3}
\end{figure}

\begin{figure}[t]                            
                                  \includegraphics[width=12 cm]
                                  {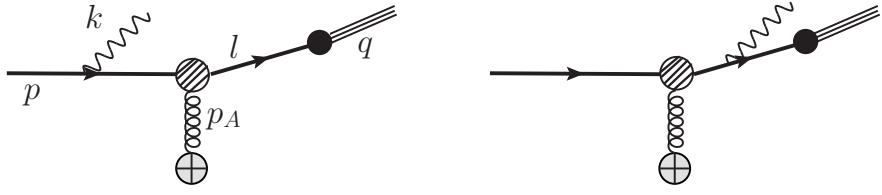} \caption{
                              The diagrams (at leading log approximation) contributing to the prompt photon+hadron production within the color glass condensate picture. The crossed white blob denotes the interaction of the projectile quark to all orders with the strong background field of the target nucleus A.  The black blob represent the quark-hadron fragmentation process.    }
\label{fig1}
\end{figure}

In this approach a fast valence quark from the projectile proton radiates a photon before and after multiply scattering on the color glass condensate target, see \fig{fig1}. In this picture, the projectile is treated in the collinear factorization, and therefore the photon radiation from quark at this level has the standard features of  pQCD, including the back-to-back correlation in the transverse momentum. As a result of multiple scatterings, the quark acquires a transverse momentum comparable with the saturation scale, the only relevant scale in the system, and the intrinsic angular correlations are washed way.

In \fig{fig1-1}, we show $P(\Delta \theta)$ at forward rapidity $\eta_h=\eta_{\gamma}=3$ for  $q_h=k_t=2$ GeV at RHIC and $q_h=k_t=6$ GeV at the LHC for two different initial saturation scale for proton $Q^2_{0p}=0.168, 0.2\,\text{GeV}^2$ and nuclei $Q_{0A}^2=3\div4 Q_{0p}^2$. For such low $p_t$'s we are most likely probing the saturation region of the nuclear wave function due to the small values of $x_g$. It is clear that the away-side prompt photon-hadron cross-section (at $\Delta \theta\approx \pi$)  is suppressed for the bigger saturation scale (corresponding to a denser system).  It is also seen from \fig{fig1-1} that $P(\Delta \theta)$ is very sensitive to the initial saturation scale. Unfortunately, this bring rather large theoretical uncertainties. However, as we will show in the following the suppression of the way-side correlations seems to be a robust feature of our results and it  less depends on our theoretical uncertainties.

In \fig{fig2} we show the relative azimuthal correlations obtained from the rcBK solution  for a fixed initial saturation scale  $Q^2_{0s}=0.168\,\text{GeV}^2$ and $Q^2_{0A}=3 Q_{0p}^2$  at forward rapidities $\eta_h=\eta_{\gamma}=3$ at RHIC $\sqrt{S}=0.2$ TeV for two different kinematics windows of transverse momenta: We show in top panel,  the results with a  fix transverse momentum of  prompt photon $q_t=5$ GeV at different transverse momentum of produced hadron, and in down panel, with a fixed transverse momentum of hadron $k_t=5$ GeV but at various transverse momentum of the produced prompt photon. It is seen, the relative azimuthal correlation is suppressed at $\Delta \theta=\pi$ as the transverse momentum of  the produced hadron or prompt photon decreases and becomes comparable to the actual saturation scale of the system. This is the case regardless which of two transverse momenta of the hadron or prompt photon decreases. In \fig{fig2} lower panel, it is seen that when the prompt photon transverse momentum becomes comparable with the saturation effect the 
away-side azimuthal angular correlation of  photon-hadron completely washes away.  The same effect happens at lower transverse momentum of the produced hadron. This is simply because of fragmentation effect namely the transverse momentum of the produced parton  (that should be compared with the saturation scale) is higher than the transverse momentum of the fragmented hadron. Again, the suppression of away-side correlations is clearly due to the saturation effect since as we lower the transverse momentum of the produced particle, the system of hadron-photon become more sensitive to the small-x gluon saturation.  
Note that the hadron-photon cross-section in \eq{qh-f}  has collinear singularity. Therefore for a proper investigation of the correlations at $\Delta \theta\approx 0$, in principle, one should first extract the collinear singularity in a same fashion as demonstrated in section II by introducing the quark-photon fragmentation function. Therefore, our results at near-side  $\Delta \theta \approx 0$ should be less reliable. Nevertheless, we expect that the sensitivity to the collinear singularity effect should drop out in the correlation defined in \eq{az} via normalization.  We checked that contrary to the away-side correlations, the near-side peak is not sensitive to the saturation physics as the correlations does not change with varying the density of the system, see  also \fig{fig1-1}.

In order to further understand the relative sensitivity of the away side peak to saturation dynamics, in \fig{fig3}, $P(\Delta \theta)$ we compare at various energies at RHIC and the LHC for a fixed transverse momentum of the produced prompt photon $k_t=5$ GeV and hadron $q_t=3$ GeV at rapidity $\eta_h=\eta_{\gamma}=3$. 
It is clear that the away side peak goes away as one increases the energy. Again this is due to the fact that as we increase the energy, the gluon density increases and non-linear gluon recombination or the saturation effect becomes important. From \fig{fig3}, it is obvious that at the LHC the away-side azimuthal correlations of photon-hadron will be strongly suppressed.

We conclude that the suppression of the away-side azimuthal photon-hadron correlations  defined via \eq{az}, with decreasing the transverse momentum of the produced prompt photon or hadron, or increasing the energy, or increasing the size/density of system, all uniquely can be explained within the universal picture of gluon saturation without invoking any new parameters or ingredients to our model.

\section{Summary} 
We have investigated prompt photon production and prompt photon-hadron
azimuthal angular correlations in proton-proton and proton-nucleus collisions using the Color
Glass Condensate formalism. We have provided predictions in the kinematic regions appropriate to
RHIC and the LHC experiments. We have shown that single inclusive and direct prompt photon production cross section in p(d)A collisions at forward rapidities at both RHIC and the LHC is suppressed, as compared to normalized production cross section in proton-proton collisions. At RHIC, a good portion of the predicted
suppression is due to the projectile being a deuteron rather than a proton. This suppression is larger at the LHC compared to RHIC which is even more impressive given that the projectile at the LHC is a proton. We showed that direct photon production is most affected by gluon saturation effects in the target nucleus than the fragmentation photons.  However, at the LHC energies at forward rapidities the nuclear modification suppression for direct, fragmentation and inclusive prompt photon production is rather similar. We showed that the nuclear modification factor $R_{p(d)A}^{\gamma}$ for inclusive prompt photon production at RHIC and the LHC is a sensitive probe of small-x dynamics.  We note that our results based on gluon saturation dynamics and using the Color Glass Condensate formalism are different from those coming from the collinear factorization \cite{pqcd-models}. Therefore, $R_{p(d)A}^{\gamma}$ measurement  at RHIC and the LHC is a crucial test of different factorization schemes, see also Refs.\,\cite{me-jamal1,me-fac,fac} for other observables.

We have also studied prompt photon-hadron azimuthal angular correlations in kinematic regions 
which can be probed by RHIC and the LHC experiments. It is shown that the away side peak
in photon-hadron angular correlation goes away as one lowers the final state particle's momenta,
very similarly to the disappearance of the away side peak in di-hadron correlations in forward
rapidity dA collisions at RHIC. At fixed transverse momenta, the suppression of the away side peak
gets stronger as one goes to larger rapidities (more forward) or higher energy or denser system as expected, due to stronger saturation
effects in the target nucleus. Presently, we are not aware of any alternative approach which leads to this novel phenomenon.  Note that in contrast to the nuclear modification factor for prompt photon, the prompt photon-hadron azimuthal angular correlation defined via Eq.\,(\ref{az}) is free from the
 isospin effect, and can be considered as a cleaner probe of saturation effect. 
Finally, we emphasize that prompt photon-hadron azimuthal angular correlations suffers from much less theoretical uncertainties as compared to di-hadron azimuthal angular correlations and thus a measurement of this correlation would go a long way toward establishing the dominance of gluon saturation effects at small $x_g$.

It will be interesting to see what the predictions of pQCD-motivated 
models~\cite{pqcd-models-2} are for photon-hadron azimuthal angular correlations. In these models one usually needs to combine models of higher twist shadowing, the Cronin effect and cold matter energy loss in order to describe the data on single inclusive hadron production and di-hadron azimuthal angular correlations. 
The advantage of the CGC formalism is that the same framework can be used to describe nuclear shadowing of 
structure functions~\cite{shad} at small $x$ and includes transverse momentum broadening (the Cronin effect)~\cite{jnv}. It does not however include cold matter energy loss due to longitudinal momentum transfer between the projectile and the target which may be important at the very forward rapidities. It is not clear at the moment how to calculate this effect from first principles QCD.  Even though this energy loss itself is small, due to steepness of the production cross section at forward rapidity, it can suppress the cross section significantly.

\appendix
\section{}

The purpose of this appendix is to define the kinematics and derive
the needed relations between various light-cone energy fractions 
which appear in the production cross sections used. This is slightly 
different from the standard relations used in production cross sections
based on collinear factorization theorems of pQCD. We first consider
scattering of a quark on the target where a photon and a 
quark are produced, depicted in Fig.~1, 
\be
q(p)\, A (p_A)\to \gamma(k) \, q(l) \, X,
\ee
where $A$ is a label for the multi-gluon state, described by a classical field representing 
a proton or nucleus target. In the standard pQCD (leading twist)
kinematics, only one parton from the target interacts. This is not the case here since
the target is described by a classical gluon field representing a multi-gluon state 
with intrinsic momentum rather than an individual gluon with a well defined energy 
fraction $x_g$ and zero transverse momentum. Nevertheless, since most of the gluons in 
the target wave function have momentum of order $Q_s$, one can think of the state describing 
the target as being labeled by a (four) momentum $p_A$. In this sense, the 
gluons in the target collectively carry fraction $x_g$ of the target energy and have 
intrinsic transverse momentum denoted $p_{A,t}$. This also means that there is no integration
over $x_g$ in our case unlike the collinearly factorized cross sections in pQCD (this basically
corresponds to setting $x_g$ equal to the lower limit of $x_g$ integration in pQCD cross sections).
We thus have 
\begin{eqnarray}\label{a1}
p^\mu&=&\left(p^-=x_q\sqrt{S/2},~p^+=0,~p_t=0 \right),\nonumber\\
P^\mu&=&\left(P^-=\sqrt{S/2},~P^+=0,~P_t=0 \right),\nonumber\\
p_A^\mu&=&\left(p^{-}_A=0,~p_A^{+}=x_g\sqrt{S/2},~p_{A,t} \right),\nonumber\\
P_A^\mu&=&\left(P^{-}_A=0,~P_A^{+}=\sqrt{S/2},~P_{A,t} = 0 \right),\nonumber\\
l^\mu&=&\left(l^-,~l^+=l_t^2/2l^-,~l_t \right),\nonumber\\
q^\mu&=&\left(q^- = z_f\, l^-,~q^+=q_t^2/2q^-,~q_t = z_f\, l_t \right),\nonumber\\
k^\mu&=& \left(k^-,~k^+=k_t^2/2k^-,~k_t \right), \
\end{eqnarray}
where $P^\mu , P_A^\mu, q^\mu$ are the momenta of the incoming projectile, target
and the produced hadron respectively. (Pseudo)-rapidities of the produced quark and 
photon are related to their energies via 
\begin{equation}\label{a2}
l^-=\frac{l_t}{\sqrt{2}}e^{\eta_h},\hspace{2cm} k^-=\frac{k_t}{\sqrt{2}}e^{\eta_\gamma},
\end{equation} 
Imposing energy-momentum conservation at the partonic level via $\delta^4 (p + p_A - l -k)$ 
and using \eq{a1} leads to
\begin{eqnarray}
p^-&=&k^- + l^-,\label{a31} \\
p^{+}_A&=&k^+ + l^+,\label{a32}\\
\vec{p}_{A,t} &=& \vec{k}_t + \vec{l}_t. 
\end{eqnarray}
The above relations and \eq{a1} (and the on mass shell condition) can be used 
to derive the following expressions for the energy fractions $x_q,x_g$. We obtain, 
\begin{eqnarray}\label{a4}
x_q&=&x_{\bar{q}}=\frac{1}{\sqrt{S}}\left(k_t\, e^{\eta_{\gamma}}+\frac{q_t}{z_f}\, e^{\eta_{h}}\right),\\
x_g&=&\frac{1}{\sqrt{S}}\left(k_t\,e^{-\eta_{\gamma}}+ \frac{q_t}{z_f}\, e^{-\eta_{h}}\right),\label{a7}\
\end{eqnarray}
where the final hadron transverse momentum and rapidity are denoted by $q_t$ and $\eta_h$, and we
used $z_f=q_t/l_t$. Note that light-cone momentum fraction $x_g$ appears in
the dipole forward scattering amplitude $N_F (b_t, r_t, x_g)$ whereas $x_q$ is the fraction of the projectile
proton (deuteron) carried by the incident quark, see \eq{a1}. To derive an expression for 
the lower limit of $z_f$ integration in \eq{qh-f}, we note that $0 \le \, x_q \, \le \, 1$. Using 
the relation between the minus components of the four momenta given above, we get
\be
x_q \, \sqrt{S/2} = k^- + \frac{q^-}{z_f}.
\ee
The minimum value of $z_f$ occurs when $x_q$ is maximum, i.e., $x_q = 1$. We then have
\be
z_{f}^{min} = {q^- \over \sqrt{S/2} - k^-},
\ee
which can be written in terms of the transverse momenta and rapidities of the final state hadron
and photon as
\be
z_{f}^{min} = \frac{q_t}{\sqrt{S}} \, {e^{\eta_h} \over 1 - \frac{k_t}{\sqrt{S}}\, e^{\eta_\gamma}}.
\ee

We now consider the kinematics of single inclusive photon production cross section. 
The cross section is obtained from \eq{cs_gen} by integrating over the final 
state quark momenta. This requires some care as we have now explicitly separated direct
and fragmentation photons in \eq{pho2}. Again using  Eqs.\,(\ref{a1}, \ref{a2}, \ref{a31}), we obtain the following relation, 
\be
x_g=\frac{1}{\sqrt{S}}\left(k_t\,e^{-\eta_{\gamma}}+ \frac{l^2_t}{\bar{l}\sqrt{2}}\right), \label{xg1}
\ee
where opposite to \eq{a7}, we avoided to introduce $\eta_h$ and $z_f$.  One can use the energy-momentum delta functions in
\eq{a31} to obtain the following relation
\be
\bar{l}=x_q\sqrt{S/2}-k_t/\sqrt{2}e^{-\eta_{\gamma}}.
\ee
Now using the above relation and \eq{xg1}, we obtain
\be\label{a5}
x_g = \frac{1}{x_q\, S} \left[{k_t^2\over z} + {l_t^2\over 1-z}\right], 
\ee
where the parameter $z$ is the fraction of energy of parton carried away by photon and it is defined as follows,
\be
z \equiv \frac{k^-}{p^-} = \frac{k_t}{x_q\, \sqrt{S}}e^{\eta_{\gamma}}. 
\ee
In case of direct photons with transverse momentum $k_t$, one should shift momentum $\vec{l_t} \rightarrow \vec{l_t} - \vec{k_t}$ in \eq{a5} 
(this is how we obtained the expersion \eq{pho2}). Assuming that $l_t<<k_t$, we get
\be\label{a6}
\bar{x}_g = \frac{1}{x_q\, S} {k_t^2 \over z (1-z)},
\ee
where we have now used $\bar{x}_g$ to denote the light cone momentum fraction of the target carried
by gluons for production of direct photons so as to distinguish it from the momentum fraction involved
in production of fragmentation photons. In the later case, the integration variable 
$l_t$ has been shifted twice. Implementing the shifts in \eq{a5} and noting that the $l_t$ 
integration in the fragmentation photon production cross section is dominated by its singularity 
at $l_t\rightarrow 0$ we get, for fragmentation photons, 
\be
x_g=\frac{k^2}{z^2 x_q S}.
\ee

\begin{acknowledgments}
We are grateful to Javier Albacete for useful communications and for providing us with the latest tables for the numerical solution of rcBK equation. We would also like to thank William Brooks and Raju Venugopalan for useful discussions. We acknowledge and thank Barbara Jacak, John Lajoie and Richard Seto for helpful discussions about PHENIX detector capabilities. J.J-M. is supported in part by the DOE Office of
Nuclear Physics through Grant No.\ DE-FG02-09ER41620, from the ``Lab
Directed Research and Development'' grant LDRD~10-043 (Brookhaven
National Laboratory), and from The City University of New York through
the PSC-CUNY Research Program, grant 64554-00 42. The work of A.H.R is
supported in part by Fondecyt grants 1110781.
\end{acknowledgments}


\end{document}